\documentclass[review]{elsarticle}
\usepackage{soul}
\usepackage{placeins}

\usepackage[hyphens]{url}

\usepackage[displaymath]{lineno}
\usepackage{hyperref}
\hypersetup{
    colorlinks=true,
    linkcolor=blue,
    filecolor=magenta,      
    urlcolor=cyan,
    pdftitle={Overleaf Example},
    pdfpagemode=FullScreen,
    }

\usepackage{siunitx}

\usepackage{graphicx,pstricks}
\usepackage{graphics}
\usepackage{epsfig}
\usepackage{amssymb}
\usepackage{amsmath, amsthm}
\usepackage{amsfonts}
\usepackage{algorithm}
\usepackage{tabu}
\usepackage{multirow}
\usepackage{subfig}
\usepackage{subcaption}
\usepackage{caption}
\usepackage{changepage}
\usepackage{algpseudocode}
\usepackage{xcolor}
\newcommand{\stateVar}{u}
\newcommand{\state}{\mathbf{\stateVar}}
\newcommand{\stateInit}{\state_{0}}

\usepackage{color}
\definecolor{green2}{rgb}{0, 0.5, 0}
\definecolor{ao(english)}{rgb}{0.0, 0.5, 0.0}
\newcommand{\YC}[1]{\textcolor{red}{#1}}

\journal{}

\begin{document}

\begin{frontmatter}

\title{GPLaSDI:  Gaussian Process-based Interpretable Latent Space Dynamics Identification through Deep Autoencoder}

\author[cee]{Christophe Bonneville\corref{correspondingauthor}}
\author[llnl]{Youngsoo Choi}
\author[llnl]{Debojyoti Ghosh}
\author[llnl]{Jonathan L. Belof}
\cortext[correspondingauthor]{Corresponding Author: \href{mailto:cpb97@cornell.edu}{\texttt{cpb97@cornell.edu}}}

\address[cee]{Cornell University, Ithaca, NY 14850, United States}
\address[llnl]{Lawrence Livermore National Laboratory, Livermore, CA 94550, United States}

\begin{abstract}
 Numerically solving partial differential equations (PDEs) can be challenging and computationally expensive. This has led to the development of reduced-order models (ROMs) that are accurate but faster than full order models (FOMs). Recently, machine learning advances have enabled the creation of non-linear projection methods, such as Latent Space Dynamics Identification (LaSDI). LaSDI maps full-order PDE solutions to a latent space using autoencoders and learns the system of ODEs governing the latent space dynamics. By interpolating and solving the ODE system in the reduced latent space, fast and accurate ROM predictions can be made by feeding the predicted latent space dynamics into the decoder. In this paper, we introduce GPLaSDI, a novel LaSDI-based framework that relies on Gaussian process (GP) for latent space ODE interpolations. Using GPs offers two significant advantages. First, it enables the quantification of uncertainty over the ROM predictions. Second, leveraging this prediction uncertainty allows for efficient adaptive training through a greedy selection of additional training data points. This approach does not require prior knowledge of the underlying PDEs. Consequently, GPLaSDI is inherently non-intrusive and can be applied to problems without a known PDE or its residual. We demonstrate the effectiveness of our approach on the Burgers equation, Vlasov equation for plasma physics, and a rising thermal bubble problem. Our proposed method achieves between 200 and 100,000 times speed-up, with up to 7$\%$ relative error.
\end{abstract}


\end{frontmatter}


\section{Introduction}

Over the past few decades, the advancement of numerical simulation techniques for understanding physical phenomena has been remarkable, leading to increased sophistication and accuracy. Simultaneously, computational hardware has undergone significant improvements, becoming more powerful and affordable. As a result, numerical simulations are now extensively utilized across various domains, including engineering design, digital twins, decision making \cite{alma991043311449403276, JONES202036, Journal, article, sep-simulations-science}, and diverse fields such as aerospace, automotive, electronics, physics, biology \cite{cummings_mason_morton_mcdaniel_2015, 6db924dfeff44d159ab577c1aefed6ef, car1, 9043275, Peterson_b1998, rylander, thijssen_2007, russel}. 

In engineering and physics, computational simulations often involve solving partial differential equations (PDEs) through different numerical techniques like finite difference/volume/element methods, particle methods, and more. While these methods have demonstrated their accuracy and capability to provide high-fidelity simulations when applied correctly, they can be computationally demanding. This is particularly evident in time-dependent multiscale problems that encompass complex physical phenomena (e.g., turbulent fluid flows, fusion device plasma dynamics, astrodynamic flows) on highly refined grids or meshes. Consequently, performing a large number of forward simulations using high-fidelity solvers can pose significant computational challenges, especially in situations like uncertainty quantification \cite{10754/656260, Smith2013UncertaintyQ, Sternfels_2013}, inverse problems \cite{10754/656260, https://doi.org/10.1002/nme.2746, https://doi.org/10.1002/2016RS005998, Sternfels_2013}, design optimization \cite{do1,do2}, and optimal control \cite{oc}.

The computational bottleneck associated with high-fidelity simulations has prompted the development of reduced-order models (ROMs). The primary objective of ROMs is to simplify the computations involved in the full-order model (i.e., the high-fidelity simulation) by reducing the order of the problem. Although ROM predictions are generally less accurate, they offer significantly faster results, making them highly appealing when a slight decrease in accuracy is tolerable. Several ROM techniques rely on the projection of snapshot data from the full-order model.

Linear projection methods such as the proper orthogonal decomposition (POD) \cite{doi:10.1146/annurev.fl.25.010193.002543}, the reduced basis method \cite{rbm}, and the balanced truncation method \cite{Safonov1988ASM} have gained popularity in ROM applications. These linear-subspace methods have been successfully applied to various equations such as the Burgers and Navier-Stokes equations \cite{https://doi.org/10.48550/arxiv.2203.16494, doi:10.1137/19M1242963, Stabile_2018, https://doi.org/10.1002/num.21835}, Lagrangian hydrodynamics \cite{Copeland_2022, https://doi.org/10.48550/arxiv.2201.07335}, advection-diffusion problems \cite{MCLAUGHLIN20162407, math9141690}, and design optimization \cite{https://doi.org/10.48550/arxiv.1909.11320, mcbanechoi}. More recently, non-linear projection methods \cite{https://doi.org/10.48550/arxiv.2009.11990, https://doi.org/10.48550/arxiv.2011.07727, LEE2020108973, diaz2023fast} utilizing autoencoders \cite{doi:10.1126/science.1127647, NIPS1992_cdc0d6e6} have emerged as an alternative and have shown superior performance in advection-dominated problems \cite{https://doi.org/10.48550/arxiv.2009.11990, lasdi, glasdi}.

Data-driven projection-based ROM methods can generally be categorized into two types: intrusive and non-intrusive. Intrusive ROMs are physics-informed models that require knowledge of the governing equation \cite{rbm,diaz2023fast,https://doi.org/10.48550/arxiv.2011.07727,LEE2020108973,doi:10.1137/19M1242963,https://doi.org/10.1002/num.21835,Copeland_2022,math9141690,mcbane2022stress}. This characteristic enhances the robustness of predictions and often necessitates less data from the full-order model (FOM). However, intrusive ROMs also demand access to the FOM solver and specific implementation details, such as the discretized residual of the PDE.

In contrast, non-intrusive ROMs are independent of the governing equation and rely solely on data-driven techniques. These methods typically utilize interpolation techniques to map parameters to their corresponding ROM predictions \cite{osti_1420279, Marjavaara2006CFDDO, unknown, kutz_2017}. However, being purely black-box approaches, these methods lack interpretability and robustness. Sometimes, they struggle to generalize accurately and may exhibit limitations in their performance.

To overcome these challenges, recent approaches have focused on combining projection methods with latent space dynamics learning. These approaches view the latent space as a dynamical system governed by a set of ordinary differential equations (ODEs). By accurately identifying these ODEs, the dynamics of the latent space can be predicted and projected back into the space of full-order solutions.

Various methods have been proposed for identifying governing equations from data \cite{koza:1994:SandC, doi:10.1126/science.1165893}, including the widely used \textit{Sparse Identification of Non-Linear Dynamics} (SINDy) \cite{doi:10.1073/pnas.1517384113}. SINDy constructs a library of terms that could potentially be part of the governing ODEs and estimates the associated coefficients using linear regression. This method has gained popularity and has been extended to a broad range of SINDy-based identification algorithms \cite{doi:10.1126/sciadv.1602614, https://doi.org/10.48550/arxiv.2211.10575, https://doi.org/10.48550/arxiv.2205.10965, doi:10.1098/rsos.211823, Messenger_2021, Chen_2021, BONNEVILLE2022100115, STEPHANY2022360, https://doi.org/10.48550/arxiv.2212.04971}.

Champion et al. \cite{doi:10.1073/pnas.1906995116} introduced an approach that combines an autoencoder with SINDy to identify sets of ODEs governing the dynamics of the latent space. While promising, the identified ODEs are not parameterized based on simulation parameters, limiting the method's generalizability. Bai and Peng \cite{https://doi.org/10.48550/arxiv.2106.09658} proposed a similar approach but with a linear projection using proper orthogonal decomposition (POD). They also introduced parameterization of the latent space ODEs, enabling ROM predictions for any point in the parameter space. However, this method exhibits limitations when applied to advection-dominated problems due to the constraints of POD.

Fries et al. \cite{lasdi} proposed a framework called the Latent Space Dynamic Identification (LaSDI), which combines autoencoders with a parametric SINDy identification of the latent space. In LaSDI, a set of ODEs corresponding to each training data point from the full-order model (FOM) is estimated. The coefficients of each ODE are then interpolated based on the FOM parameters using radial basis functions (RBF). However, the sequential training of the autoencoder and SINDy identifications in LaSDI can sometimes lead to a complex latent space with poorly conditioned and/or inaccurate sets of governing ODEs. Additionally, the training FOM dataset in LaSDI is generated on a predefined grid in the parameter space, which may not be optimal. The uniform grid may result in insufficient data in certain regions of the parameter space and excessive data in others, affecting the accuracy of the model.

To address these issues, He et al. \cite{glasdi} introduced the greedy-LaSDI (gLaSDI) framework. In gLaSDI, the autoencoder and SINDy latent space identifications are trained simultaneously, and the FOM data points are sampled sequentially during training. The training starts with only a few points, and at a fixed sampling rate, the model's prediction accuracy is evaluated by plugging the decoder output into the PDE residual. The parameter yielding the largest residual error is selected as the new sampling point, and a FOM simulation is performed to obtain the solution, which is then added to the training dataset. This iterative process ensures that the model focuses on regions of the parameter space where accuracy is needed. Although gLaSDI is robust and accurate, it inherently requires knowledge of the PDE residual, making it an intrusive ROM method. Therefore, gLaSDI may not be suitable for problems where the residual is unknown, difficult to implement, or computationally expensive to evaluate.

In this paper, we present the (greedy) Gaussian-Process Latent Space Identification (GPLaSDI), a non-intrusive extension of the greedy LaSDI framework. One of the main sources of error in LaSDI and gLaSDI arises from potential inaccuracies in the interpolation of the ODE coefficients. To address this issue, we propose replacing deterministic interpolation methods (such as RBF interpolation in LaSDI \cite{lasdi} and $k$-NN convex interpolation in gLaSDI \cite{glasdi}) with a Bayesian interpolation technique called Gaussian Process (GP) \cite{books/lib/RasmussenW06}.

A GP can provide confidence intervals for its predictions, allowing us to quantify the uncertainty in the sets of ODEs governing the latent space. This uncertainty can then be propagated to the decoder, enabling the generation of ROM prediction confidence intervals for any test point in the parameter space. We adopt a sequential procedure for sampling additional FOM training data, similar to gLaSDI. However, instead of relying on the PDE residual to select the next parameter for sampling, we choose the parameter that yields the highest predictive uncertainty.

This approach offers two significant advantages. First, it makes our framework fully independent of the specific PDE and its residual, making it applicable to a wide range of problems. Second, our method can generate meaningful confidence intervals for ROM predictions, providing valuable information for assessing the reliability of the simulations.

\begin{figure}[!h]
\hspace{-3cm}
    \includegraphics[width=1.5\textwidth]{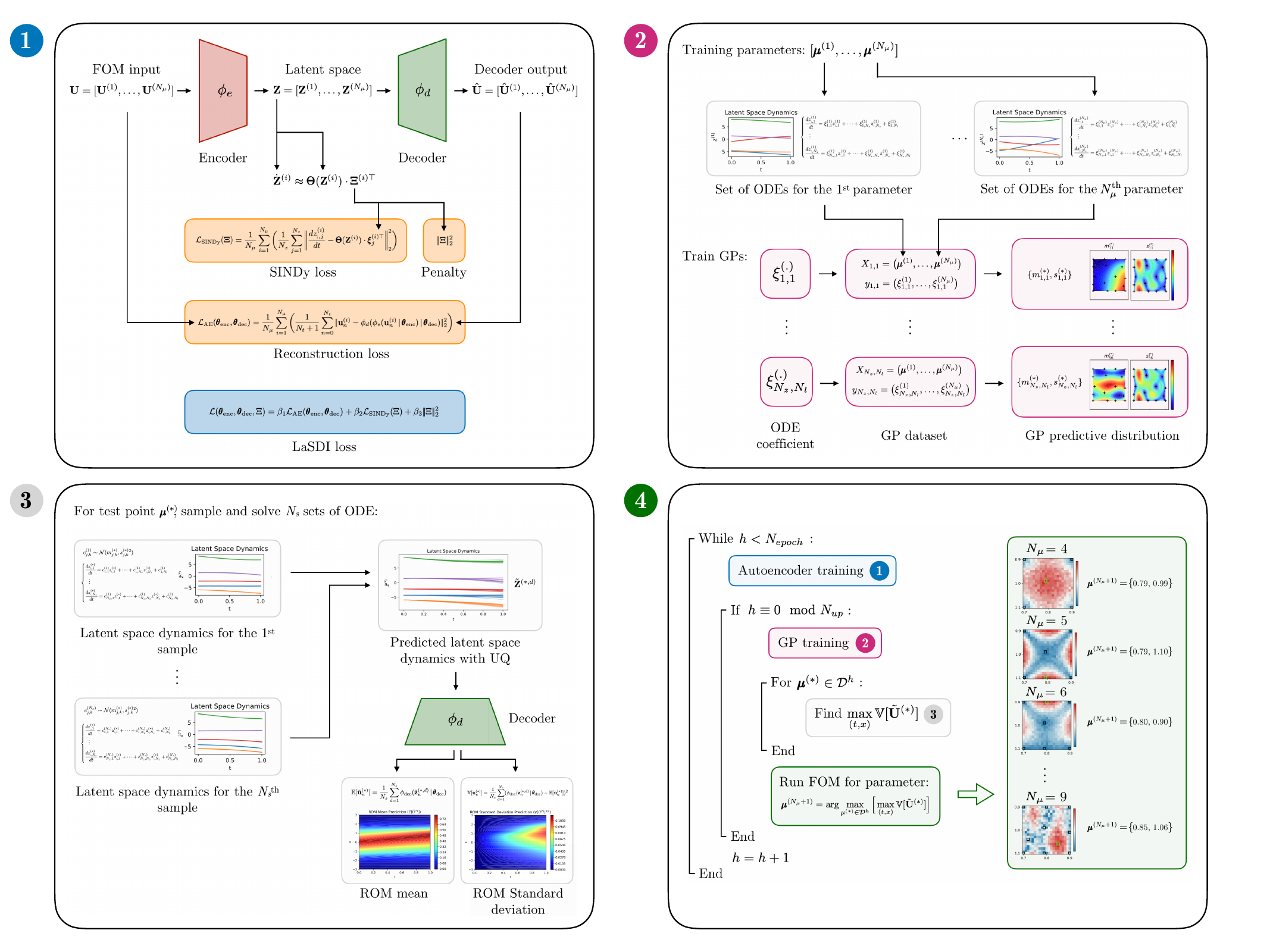}

    \caption{GPLaSDI general framework. (1) Combined training of the autoencoder and the SINDy latent space identification. (2) Interpolation of the latent space governing sets of ODEs using Gaussian Processes. (3) ROM Prediction methodology using GPLaSDI. (4) FOM training data greedy sampling algorithm, using (1), (2) and (3). 
    }
    \label{framework}
\end{figure}

The details of the GPLaSDI framework are presented in Section \ref{gplasdi}, where we introduce the utilization of autoencoders in Section \ref{autoencoder}, the application of SINDy in Section \ref{sindy}, the adoption of GP interpolation in Section \ref{gp}, and the incorporation of variance-based greedy sampling in Section \ref{greedy}. The overall structure of the GPLaSDI framework is summarized in Figure \ref{framework}, and the algorithmic procedure is outlined in Algorithm \ref{alg:cap}. To demonstrate the effectiveness of GPLaSDI, we provide case studies in Section \ref{example}. Specifically, we investigate the application of GPLaSDI to the 1D Burgers equation in Section \ref{burgers1}, the 2D Burgers equation in Section \ref{burgers2}, the 1D1V Vlasov equation for plasma physics in Section \ref{vlasov}, and the 2D rising thermal bubble problem in Section \ref{2D Rising Thermal Bubble}. Through these examples, we highlight the performance and capabilities of the GPLaSDI framework. In conclusion, we summarize the key findings and contributions of the paper in Section \ref{sec:conclusion}, providing a comprehensive overview of the work conducted in this study.

\section{GPLaSDI Framework}
\label{gplasdi}
\subsection{Governing Equation of Physical Systems}
\noindent In the following sections, we consider physical phenomenons described by governing PDEs with the following form:
\begin{equation}
\label{pde}
\begin{cases}
    \,\displaystyle\frac{\partial \state}{\partial t}=\mathbf{f}(\state,t,x\,|\,\pmb{\mu})\hspace{0.75in}(t,x)\in[0,t_{\text{max}}]\times\Omega\\[5pt]
    \,\state(t=0,x\,|\,\pmb{\mu})=\stateInit(x\,|\,\pmb{\mu})\hspace{0.35in}\pmb{\mu}\in\mathcal{D}\\
\end{cases}
\end{equation}
\noindent In Equation \eqref{pde}, the solution $\mathbf{u}$ may either represent a scalar or vector field, defined over the time--space domain $[0,t_{\text{max}}]\times\Omega$. The spatial domain $\Omega$ may be of any dimensions, and the differential operator $\mathbf{f}$ may contain linear and/or non--linear combinations of spatial derivatives and source terms. The governing equations and their initial/boundary conditions are parameterized by a parameter vector $\pmb{\mu}$. The parameter space is defined as $\mathcal{D}\subseteq\mathbb{R}^{n_\mu}$ and may take any dimension (in the following sections, we consider a 2D-parameter space, i.e. $n_\mu=2$). For a given parameter vector $\pmb{\mu}^{(i)}\in\mathcal{D}$, we assume in the following sections that we have access to the corresponding discretized solution of Equation \eqref{pde}, denoted as $\mathbf{U}^{(i)}$. $\mathbf{U}^{(i)}$ is a matrix of concatenated snapshots $\mathbf{u}_n^{(i)}$ at each time step $n$, such that $\mathbf{U}^{(i)}=[\mathbf{u}_0^{(i)},\dots,\mathbf{u}_{N_t}^{(i)}]^\top\in\mathbb{R}^{(N_t+1)\times N_u}$. This solution is obtained using either a full order model solver or an experiment. In the following sections, every FOM solutions are organized into a $3^\text{rd}$ order tensor dataset $\mathbf{U}\in\mathbb{R}^{N_\mu \times (N_t+1)\times N_u}$ of the form:

\begin{equation}
    \mathbf{U}=[\mathbf{U}^{(1)},\dots,\mathbf{U}^{(N_\mu)}]_{N_\mu\times (N_t+1)\times N_u}
\end{equation}
\noindent $N_t$ and $N_u$ are the number of time steps and degrees of freedom, respectively, and $N_\mu$ is the number of available FOM solutions. The notations employed in this paper are summarized in Table \ref{notations}.

\begin{table}
\vspace{-0.5in}
\hspace{-0.7in}
    \begin{tabular}{l|l}
        Notation & Description \\\hline
         $\mathbf{u}_n^{(i)}$ & FOM snapshot at time step $n$ and parameter $\pmb{\mu}^{(i)}$\\
         $\hat{\mathbf{u}}_n^{(i)}$ & Reconstructed autoencoder FOM snapshot at time step $n$ and parameter $\pmb{\mu}^{(i)}$\\
         $\tilde{\mathbf{u}}_n^{(*,d)}$ & Predicted ROM solution at time step $n$ and test parameter $\pmb{\mu}^{(*)}$\\
         $\mathbf{z}_n^{(i)}$ & Latent space representation of snapshot $\mathbf{u}_n^{(i)}$\\
         $\tilde{\mathbf{z}}_n^{(*,d)}$ & Predicted latent space dynamics at time step $n$, for test parameter $\pmb{\mu}^{(*)}$ and sample $d$\\
         $\pmb{\mu}^{(i)}$/$\pmb{\mu}^{(*)}$ & Training parameter/Test parameter\\
         $\phi_e$/$\phi_d$ & Encoder/Decoder\\
         $\pmb{\theta}_\text{enc}$/$\pmb{\theta}_\text{dec}$ & Encoder weights/Decoder weights\\
         $\pmb{\Xi}^{(i)}=\{\xi^{(i)}_{j,k}\}$ &  SINDy coefficients for the set of ODEs associated to parameter $\pmb{\mu}^{(i)}$\\
         $\pmb{\theta}_\text{gp}=\{\pmb{\theta}^{j,k}_\text{gp}\}$ & Sets of GP hyperparameters\\
         $\pmb{\Theta}(\cdot)$ & SINDy dictionary \\
         $\mathcal{D}$/$\mathcal{D}^h$ & Parameter space/Discretized parameter space\\
         $[\![1,N]\!]$ & a set of integers, i.e., $\{1,\ldots,N\}$\\
         $n_\mu$ & Dimension of $\mathcal{D}$\\
         $N_u$ & Number of degrees of freedom in the FOM solution\\
         $N_t$ & Number of time steps in the FOM solution\\
         $N_\mu$ & Number of parameters in the training dataset\\
         $N_z$ & Number of variables in the latent space\\
         $N_l$ & Number of SINDy candidates\\
         $N_s$ & Number of samples taken from each GP predictive distribution\\
         $N_{epoch}$ & Number of training epochs\\
         $N_{up}$ & Greedy sampling rate\\
         $\beta_1$/$\beta_2$/$\beta_3$ & Loss hyperparameters\\
         $m^{(*)}_{j,k}$/$s^{(*)}_{j,k}$ & Predictive mean/standard deviation of each GP\\
         $c^{(d)}_{j,k}$ & Sample $d$ from $\mathcal{N}(m^{(*)}_{j,k}/s^{(*)2}_{j,k})$\\
         $n$ & Dummy variable for time step indexing\\
         $i$ & Dummy variable for parameter indexing \\
         $j$ & Dummy variable for latent variable indexing\\
         $k$ & Dummy variable for SINDy candidate indexing\\
         $d$ & Dummy variable for GP sampling indexing\\
         $h$ & Dummy variable for epoch count\\
    \end{tabular}
    \caption{Table of notations}
    \label{notations}
\end{table}

\subsection{Autoencoders}
\label{autoencoder}




\noindent  An autoencoder \cite{doi:10.1126/science.1127647, Goodfellow-et-al-2016} is a neural network designed specifically for compressing large datasets by reducing the dimensions of the data through nonlinear transformation. It consists of two stacked neural networks: the encoder, denoted as $\phi_e$ and parameterized by $\pmb{\theta}_\text{enc}$, and the decoder, denoted as $\phi_d$ and parameterized by $\pmb{\theta}_\text{dec}$. The encoder takes an input data snapshot $\mathbf{u}_n^{(i)}\in\mathbb{R}^{N_u}$ and produces a compressed representation $\mathbf{z}_n^{(i)}\in\mathbb{R}^{N_z}$ in a latent space. Here, $N_z$ represents the number of latent space variables, which is an arbitrary design choice, but typically chosen such that $N_z<\!\!<N_u$. Similar to the matrix $\mathbf{U}^{(i)}$, we concatenate the latent representations at each time step into a matrix $\mathbf{Z}^{(i)}=[\mathbf{z}_{0}^{(i)},\dots,\mathbf{z}_{N_t}^{(i)}]^\top\in\mathbb{R}^{(N_t+1)\times N_z}$. The latent variables for each time step and each parameter $\pmb{\mu}^{(i)}$ are stored in a third-order tensor $\mathbf{Z}\in\mathbb{R}^{N_\mu\times (N_t+1)\times N_z}$, analogous to the tensor $\mathbf{U}$, and have the following form:
\begin{equation}
    \mathbf{Z}=[\mathbf{Z}^{(1)},\dots,\mathbf{Z}^{(N_\mu)}]_{N_\mu\times (N_t+1)\times N_z}
\end{equation}
The decoder takes each $\mathbf{z}_n^{(i)}$ as input and generates a reconstructed version of $\mathbf{u}_n^{(i)}$, denoted as $\hat{\mathbf{u}}_n^{(i)}$.
\begin{equation}
\label{ae}
\begin{aligned}
    &\mathbf{z}_n^{(i)}=\phi_e(\mathbf{u}_n^{(i)}\,|\,\pmb{\theta}_\text{enc})\\
    &\hat{\mathbf{u}}_n^{(i)}=\phi_d(\mathbf{z}_n^{(i)}\,|\,\pmb{\theta}_\text{dec})
\end{aligned}
\end{equation}
\noindent The parameters of the autoencoder, denoted as $\pmb{\theta}_\text{enc}$ and $\pmb{\theta}_\text{dec}$, are learned using a numerical optimization algorithm that minimizes the $L_2$ norm of the difference between the set of input solutions $\mathbf{U}$ and the set of reconstructed solutions $\hat{\mathbf{U}}$. This is achieved by defining the reconstruction loss as follows:
\begin{equation}
\begin{aligned}
    \mathcal{L}_\text{AE}(\pmb{\theta}_\text{enc},\pmb{\theta}_\text{dec})&=|\!|\mathbf{U}-\hat{\mathbf{U}}|\!|_2^2\\
    &=\frac{1}{N_\mu}\sum_{i=1}^{N_\mu}\bigg(\frac{1}{N_t+1}\sum_{n=0}^{N_t}|\!|\mathbf{u}_n^{(i)}-\phi_d(\phi_e(\mathbf{u}_n^{(i)}\,|\,\pmb{\theta}_\text{enc})\,|\,\pmb{\theta}_\text{dec})|\!|_2^2\bigg)
\end{aligned}
\end{equation}

\subsection{Identification of Latent Space Dynamics}
\label{sindy}

\noindent The encoder performs compression of the high-dimensional physical data, e.g., a solution of PDEs, defined over space and time, into a reduced set of discrete latent variables that are defined solely over time. Consequently, the latent space can be regarded as a dynamical system governed by sets of ordinary differential equations (ODEs). This characteristic forms a fundamental aspect of LaSDI (Latent Space Dynamics Identification) algorithms, which enable the compression of dynamical systems governed by PDEs into systems governed by ODEs \cite{lasdi}. At each time step, the dynamics of the latent variables in the latent space can be described by an equation of the following form:
\begin{equation}
    \frac{d\mathbf{z}_n^{(i)}}{dt}=\psi_{DI}(\mathbf{z}_n^{(i)}\,|\,\pmb{\mu}^{(i)})\hspace{0.3in}(i,n)\in[\![1,N_\mu]\!]\times[\![0,N_t]\!]
\end{equation}
This equation can be unified across all time steps as:
\begin{equation}
\label{ode}
    \dot{\mathbf{Z}}^{(i)}=\psi_{DI}(\mathbf{Z}^{(i)}\,|\,\pmb{\mu}^{(i)})
\end{equation}
The system of ODEs governing the dynamics of the latent space can be determined using a technique called SINDy (Sparse Identification of Nonlinear Dynamics) \cite{doi:10.1073/pnas.1517384113}. In SINDy, a dictionary $\pmb{\Theta}(\mathbf{Z}^{(i)})\in\mathbb{R}^{(N_t+1)\times N_l}$ is constructed, consisting of $N_l$ linear and nonlinear candidate terms that may be involved in the set of ODEs. The approach assumes that the time derivatives $\dot{\mathbf{Z}}$ can be expressed as a linear combination of these candidate terms. Hence, the right-hand side of Equation \eqref{ode} can be approximated as follows:
\begin{equation}
    \dot{\mathbf{Z}}^{(i)}\approx
    \pmb{\Theta}(\mathbf{Z}^{(i)})\cdot\pmb{\Xi}^{(i)\top}
\end{equation}
where $\pmb{\Xi}^{(i)}\in\mathbb{R}^{N_z\times N_l}$ denotes a coefficient matrix. The selection of terms in $\pmb{\Theta}(\cdot)$ is arbitrary. Including a broader range of SINDy terms may potentially capture the latent dynamics more accurately, but it can also result in sets of ODEs that are more challenging to solve numerically. In the subsequent sections, we will typically limit the dictionary to constant and linear terms. Interestingly, we have found that this restricted choice of terms is generally adequate to achieve satisfactory performance:
\begin{equation}
    \pmb{\Theta}(\mathbf{Z}^{(i)})=
    \begin{bmatrix}
    1 & \mathbf{z}_0^{(i)}\\
    \vdots & \vdots \\
    1 & \mathbf{z}_{N_t}^{(i)}
    \end{bmatrix}
    =
    \begin{bmatrix}
    1 & z_{0,1}^{(i)} & \dots & z_{0,j}^{(i)} & \dots & z_{0,N_z}^{(i)}\\
    \vdots & \vdots & & \vdots & & \vdots \\
    1 & z_{N_t,1}^{(i)} & \dots & z_{N_t,j}^{(i)} & \dots & z_{N_t,N_z}^{(i)}
    \end{bmatrix}_{(N_t+1) \times N_l}
\end{equation}
In the system, $z^{(i)}_{n,j}$ represents the $j^\text{th}$ latent variable for parameter $\pmb{\mu}^{(i)}$ at time step $n$, such that $\mathbf{z}_n^{(i)}=[z_{n,1}^{(i)},\dots,z_{n,j}^{(i)},\dots,z_{n,N_z}^{(i)}]$. Sparse linear regressions are conducted between $\pmb{\Theta}(\mathbf{Z}^{(i)})$ and $dz^{(i)}_{:,j}/dt$ for each $j\in[\![1,N_z]\!]$. The resulting coefficients associated with the SINDy terms are stored in a vector $\pmb{\xi}_j^{(i)}=[\xi_{j,1}^{(i)},\dots,\xi_{j,N_l}^{(i)}]\in\mathbb{R}^{N_l}$. More formally, the system of SINDy regressions is expressed as:
\begin{equation}
    \frac{dz^{(i)}_{:,j}}{dt}=\begin{bmatrix}
    \dot{z}_{0,j}^{(i)}\\
    \vdots\\
    \dot{z}_{N_t,j}^{(i)}
    \end{bmatrix}
    =\pmb{\Theta}(\mathbf{Z}^{(i)})\cdot\pmb{\xi}_j^{(i)\top}
\end{equation}
Each $\pmb{\xi}_j^{(i)}$ is concatenated into a coefficient matrix $\pmb{\Xi}^{(i)}=[\pmb{\xi}_1^{(i)},\dots,\pmb{\xi}_{N_z}^{(i)}]^\top\in\mathbb{R}^{N_z\times N_l}$. In this study, the time derivatives $\dot{z}_{n,j}^{(i)}$ are estimated using a first-order finite difference with a time step $\Delta t$ that matches the time step used in the full order model (FOM) data. As there is a distinct set of ODEs governing the dynamics of the latent space for each parameter $\pmb{\mu}^{(i)}\in\mathcal{D}$, multiple SINDy regressions are performed concurrently to obtain the corresponding sets of ODE coefficients $\pmb{\Xi}^{(i)}$, where $i\in[\![1,N_\mu]\!]$. The collection of ODE coefficient matrices $\mathbf{\Xi}=[\mathbf{\Xi}^{(1)},\dots,\mathbf{\Xi}^{(N_\mu)}]\in\mathbb{R}^{N_\mu \times N_z\times N_l}$, associated with each system of ODEs, is determined by minimizing the following mean-squared-error SINDy loss:
\begin{equation}
\begin{aligned}
    \mathcal{L}_\text{SINDy}(\mathbf{\Xi})&=|\!|\dot{\mathbf{Z}}-\dot{\hat{\mathbf{Z}}}|\!|_2^2\\
    &=\frac{1}{N_\mu}\sum_{i=1}^{N_\mu}\bigg(\frac{1}{N_z}\sum_{j=1}^{N_z}\bigg|\!\bigg|\frac{dz_{:,j}^{(i)}}{dt}-\pmb{\Theta}(\mathbf{Z}^{(i)})\cdot\pmb{\xi}_j^{(i)\top}\bigg|\!\bigg|_2^2\bigg)
\end{aligned}
\end{equation}
In the LaSDI framework, the autoencoder and the SINDy sparse regressions are jointly trained using a single loss function. To prevent extreme values for the SINDy coefficients, which can result in ill-conditioned sets of ODEs, a penalty term is incorporated into the loss. Thus, the LaSDI loss function is defined as follows:
\begin{equation}
\label{loss}
    \mathcal{L}(\pmb{\theta}_\text{enc},\pmb{\theta}_\text{dec},\pmb{\Xi})=\beta_1\mathcal{L}_\text{AE}(\pmb{\theta}_\text{enc},\pmb{\theta}_\text{dec})+\beta_2\mathcal{L}_\text{SINDy}(\pmb{\Xi})+\beta_3|\!|\pmb{\Xi}|\!|_2^2,
\end{equation}
where $\beta_1$, $\beta_2$, and $\beta_3$ are weighting hyperparameters. Note that another reconstruction term to the loss function can be added, namely the $L_2$ norm between the velocity $\pmb{\dot{U}}$ and the reconstructed velocity $\pmb{\dot{\hat{U}}}$ for additional stability and accuracy. However, doing so requires to have access to $\pmb{\dot{U}}$, and computing $\pmb{\dot{\hat{U}}}$ requires to backpropagate the autoencoder a second time \cite{glasdi}. This can be expensive, especially with deeper autoencoders. Consequently in this paper, we do not use such loss term and only rely on the penalty term to maintain stability of the SINDy coefficients. As shown in the example sections, this is sufficient to obtain satisfactory accuracy. The autoencoder and SINDy coefficients can be simultaneously trained by minimizing Equation \eqref{loss} using a gradient-descent-based optimizer. In the following sections, we utilize the Adam optimizer \cite{adam} with a learning rate of $\alpha$.

\subsection{Parameterization through Coefficient Interpolation}
\label{gp}
\noindent The autoencoder learns a mapping to and from the latent space and identifies the set of ODEs governing the latent dynamics, but it does so only for each parameter $\pmb{\mu}_i\in\mathcal{D}$ associated with each training data point $\mathbf{U}^{(i)}$ (with $i\in[\![1,N_\mu]\!]$). To make prediction for any new parameter $\pmb{\mu}^{(*)}$, the system of ODEs associated to this later parameter value needs to be estimated. This can be done by finding a mapping $f:\pmb{\mu}^{(*)}\mapsto\pmb{\Xi}^{(*)}$, where $(j,k)^\text{th}$ element of $\pmb{\Xi}^{(*)}$ is denoted as $\{\xi_{j,k}^{(*)}\}_{(j,k)\in[\![1,N_z]\!]\times [\![1,N_l]\!]}$. In LaSDI and gLaSDI, $f$ is estimated through RBF interpolation \cite{lasdi} or $k-$NN convex interpolation \cite{glasdi} of each pair of data $(\pmb{\mu}^{(i)},\{\xi_{j, k}^{(i)}\})_{i\in[\![1,N_\mu]\!]}$. In this paper, we introduce a replacement for the deterministic interpolation methods, utilizing GP regression \cite{books/lib/RasmussenW06}. The utilization of GPs provides three significant advantages, which will be elaborated in the subsequent subsections:
\begin{itemize}
    \item GPs have the inherent capability to automatically quantify interpolation uncertainties by generating confidence intervals for their predictions.
    \item GPs are known for their robustness to noise, making them less prone to overfitting and mitigating the risk of incorporating incorrect SINDy coefficients.
    \item Being part of the family of Bayesian methods, GPs allow us to incorporate prior knowledge about the latent space behavior, particularly in terms of smoothness. By specifying a prior distribution, we can provide useful hints that enhance interpolation accuracy within the GP framework.
\end{itemize}
The subsequent two subsections cover essential background information on Gaussian processes (GPs) and outline their specific utilization within the context of this paper. In the first subsection, we provide an overview of GPs, while the second subsection focuses on detailing the application of GPs in our approach.

\subsubsection{Gaussian Processes}

\noindent Let's consider a dataset consisting of $N_\mu$ input--output pairs in the form of $\mathcal{S}=(X,y)$. Here, $X$ represents a set of input vectors, and $y$ represents the corresponding continuous scalar output. We assume that the output $y$ may be affected by Gaussian noise with a variance of $\sigma^2$. Our objective is to find a mapping function $f$ such that:
\begin{equation}
\label{model}
    y=f(X)+\epsilon\hspace{0.3in}\epsilon\sim\mathcal{N}(0,\sigma^2 I_{N_\mu})
\end{equation}
In the Gaussian process (GP) paradigm, the function $f$ is assumed to be drawn from a Gaussian prior distribution (Equation \eqref{prior}) \cite{books/lib/RasmussenW06}. In the literature, it is common to set the mean of the prior to 0, while the covariance is determined by a kernel function $k$ with parameters $\pmb{\theta}_\text{gp}$. The choice of kernel is arbitrary, and in this context, we adopt the radial basis function (RBF) kernel (Equation \eqref{rbf}). This choice is particularly suitable because the RBF kernel can effectively approximate any $\mathcal{C}^\infty$ function. Furthermore, there is no a priori reason to assume that the space of ordinary differential equation (ODE) coefficients lacks smoothness.
\begin{equation}
\label{prior}
    p(f\,|\,X)=\mathcal{N}(f\,|\,0,k(X,X\,|\,\pmb{\theta}_\text{gp}))
\end{equation}

\begin{equation}
    \label{rbf}
    k(X,X\,|\,\pmb{\theta}_\text{gp})=\gamma\exp\bigg(-\frac{|\!|X-X^\top|\!|_2^2}{2\lambda^2}\bigg)\hspace{0.3in}\pmb{\theta}_\text{gp}=\{\gamma,\lambda\}
\end{equation}
The likelihood (Equation \eqref{lkl}) is obtained as a direct consequence of Equation \eqref{model} and by applying Bayes' rule in combination with the prior distribution. This allows us to derive the posterior distribution over $f$ (Equation \eqref{posterior}).
\begin{equation}
\label{lkl}
    p(y\,|\,f,X)=\mathcal{N}(y\,|\,f(X),\sigma^2I_{N_\mu})
\end{equation}
\begin{equation}
\label{posterior}
    p(f\,|\,X,y)=\frac{p(y\,|\,f,X)p(f\,|\,X)}{p(y\,|\,X)}
\end{equation}
The denominator in Equation \eqref{posterior}, often referred to as the marginal likelihood, plays a crucial role in Bayesian inference. Typically, the hyperparameters $\pmb{\theta}_\text{gp}$ are selected to maximize this marginal likelihood. Alternatively, it is common to minimize the negative log--marginal--likelihood in Equation \eqref{mll}, which is equivalent but computationally more efficient in practice.
\begin{equation}
    p(y\,|\,X)=\int p(y\,|\,f,X)p(f\,|\,X)df
\end{equation}
\begin{equation}
\label{mll}
    \mathcal{L}_\text{GP}(\pmb{\theta}_\text{gp})=-\log(p(y\,|\,X))
\end{equation}
Unlike parametric Bayesian machine learning models such as Bayesian neural networks, where the posterior is defined over the model parameter space, Gaussian process models define the posterior over the function space. Consequently, predicting outputs for test input points is not as straightforward. To derive the distribution over the predictive output $y^{(*)}$ for a test input $x^{(*)}$, we need to apply both the sum rule and the product rule in combination:
\begin{equation}
\label{pred}
    p(y^{(*)}\,|\,X,y,x^{(*)})=\int p(y^{(*)}\,|\,f,x^{(*)})p(f\,|\,X,y)df
\end{equation}
In the present scenario, Bayesian inference proves to be highly tractable due to the Gaussian nature of the posterior distribution \cite{books/lib/RasmussenW06}. This implies that both the posterior distribution and Equation \eqref{pred} can be computed analytically, resulting in Gaussian distributions for both:
\begin{equation}
    p(y^{(*)}\,|\,X,y,x^{(*)})=\mathcal{N}(y^{(*)}\,|\,m^{(*)},s^{(*)2})
\end{equation}
\begin{equation}
    \begin{cases}
    \,\,m^{(*)}=k(x^{(*)},X\,|\,\pmb{\theta}_\text{gp})(k(X,X\,|\,\pmb{\theta}_\text{gp})+\sigma^2I_{N_\mu})^{-1}y\\
    \,\,s^{(*)2}=k(x^{(*)},x^{(*)}\,|\,\pmb{\theta}_\text{gp})-k(x^{(*)},X\,|\,\pmb{\theta}_\text{gp})(k(X,X\,|\,\pmb{\theta}_\text{gp})+\sigma^2I_{N_\mu})^{-1}k(X,x^{(*)}\,|\,\pmb{\theta}_\text{gp})
    \end{cases}
\end{equation}

\subsubsection{GP interpolation of the SINDy coefficients}

\noindent After training the autoencoder and obtaining the SINDy coefficients, we proceed to construct $N_{gp}=N_z\times N_l$ regression datasets, where each dataset corresponds to an ODE coefficient and consists of $N_\mu$ data points: $(X_{j,k}^{\YC{(i)}},y_{j,k}^{\YC{(i)}})=(\pmb{\mu}^{(i)},\{\xi_{j, k}^{(i)}\})_{i\in[\![0,N_\mu]\!]}$. Subsequently, a Gaussian Process (GP) is trained for each dataset. For a given test parameter $\pmb{\mu}^{(*)}$, the predictive mean and standard deviation are denoted as $m^{(*)}_{j,k}$ and $s^{(*)}_{j,k}$, respectively. To illustrate, let's consider a LaSDI system that solely incorporates constant and linear candidate terms (i.e., $N_l=N_z+1$). The system of ODEs, considering a 1-standard deviation uncertainty for the test parameter $\pmb{\mu}^{(*)}$, is as follows:
\begin{equation}
\label{gp_pred1}
    \begin{cases}
    \displaystyle\frac{dz^{(*)}_{:, 1}}{dt}=\Big(m_{1,1}^{(*)}\pm s_{1,1}^{(*)}\Big)z^{(*)}_{:, 1}+\cdots+\Big(m_{1,N_z}^{(*)}\pm s_{1,N_z}^{(*)}\Big)z^{(*)}_{:, N_z}+\Big(m_{1,N_l}^{(*)}\pm s_{1,N_l}^{(*)}\Big)\\[5pt]
    \,\,\,\,\,\vdots\\[5pt]
    \displaystyle\frac{dz^{(*)}_{:, N_z}}{dt}=\Big(m_{N_z,1}^{(*)}\pm s_{N_z,1}^{(*)}\Big)z^{(*)}_{:, 1}+\cdots+\Big(m_{N_z,N_z}^{(*)}\pm s_{N_z,N_z}^{(*)}\Big)z^{(*)}_{:, N_z}+\Big(m_{N_z,N_l}^{(*)}\pm s_{N_z,N_l}^{(*)}\Big)\\[5pt]
    \end{cases}
\end{equation}
with:
\begin{equation}
\label{gp_pred2}
    \begin{cases}
    \,\,m^{(*)}_{j,k}=k(\pmb{\mu}^{(*)},X_{j,k}\,|\,\pmb{\theta}_\text{gp}^{j,k})(k(X_{j,k},X_{j,k}\,|\,\pmb{\theta}_\text{gp}^{j,k})+\sigma^2I_{N_\mu})^{-1}y_{j,k}\\
    \,\,s^{(*)2}_{j,k}=k(\pmb{\mu}^{(*)},\pmb{\mu}^{(*)}\,|\,\pmb{\theta}_\text{gp}^{j,k})-k(\pmb{\mu}^{(*)},X_{j,k}\,|\,\pmb{\theta}_\text{gp}^{j,k})(k(X_{j,k},X_{j,k}\,|\,\pmb{\theta}_\text{gp}^{j,k})+\sigma^2I_{N_\mu})^{-1}k(X_{j,k},\pmb{\mu}^{(*)}\,|\,\pmb{\theta}_\text{gp}^{j,k})
    \end{cases}
\end{equation}


\noindent GP training typically faces scalability challenges when dealing with large datasets due to the need to invert the $N_\mu \times N_\mu$ kernel matrix, which has a computational complexity of $\mathcal{O}(N_\mu^3)$. While the computational cost of training the GPs should only become an issue which larger values of $N_\mu$ (e.g. $N_\mu>>500$), GP scalability can be addressed in different ways. One approach is to partition the parameter space into multiple subdomains and construct a separate GP for each subdomain \cite{FUHG2022114217}. This strategy effectively mitigates the computational burden associated with GPs when dealing with large datasets. Another approach is to use approximate scalable GP methods such as kernel interpolation \cite{wilson2015kernel, wilson2015thoughts} or hyperparameter cross-validation \cite{muyskens2021muygps}.


\subsection{Predicting Solutions}

\noindent
By employing the GP interpolation approach discussed in the preceding sub-section, we are able to make predictions for the governing equations of the latent space dynamics, along with associated uncertainty, for any test point $\pmb{\mu}^{(*)}\in\mathcal{D}$ within the parameter space. Consequently, we can now proceed with ease to generate Reduced Order Model (ROM) predictions using the following steps:
\begin{itemize}
    \item Compute the set of $\{m^{(*)}_{j,k},s^{(*)}_{j,k}\}$ for each $(j,k)\in[\![1,N_z]\!]\times[\![1,N_l]\!]$ and determine the governing Ordinary Differential Equaitons (ODEs), utilizing Equation \eqref{gp_pred1} and Equation \eqref{gp_pred2}, respectively.
    \item Convert the initial conditions $\mathbf{u}_0^{(*)}$ into latent space initial conditions $\mathbf{z}_0^{(*)}$ by performing a forward pass through the encoder network:
    \begin{equation}
        \mathbf{z}_0^{(*)}=\phi_\text{enc}(\mathbf{u}_0^{(*)}\,|\,\pmb{\theta}_\text{enc})
    \end{equation}
    \item Perform the simulation of the latent space dynamics by numerically solving the system of ODEs using commonly used integration schemes such as backward Euler, Runge-Kutta, or other suitable methods. The resulting simulated latent space dynamics is denoted as $\Tilde{\mathbf{Z}}^{(*)}=[\mathbf{z}_0^{(*)},\Tilde{\mathbf{z}}_1^{(*)},\dots,\Tilde{\mathbf{z}}^{(*)}_{N_t}]\in\mathbb{R}^{N_z\times (N_t+1)}$, where $N_t$ represents the number of time steps.
    \item Utilize the forward pass of $\Tilde{\mathbf{Z}}^{(*)}$ through the decoder network to generate predictions of the Full Order Model (FOM) physics. The output of the decoder is denoted as $\Tilde{\mathbf{U}}^{(*)}=[\Tilde{\mathbf{u}}_0^{(*)},\dots,\Tilde{\mathbf{u}}^{(*)}_{N_t}]\in\mathbb{R}^{N_u\times (N_t+1)}$.
\end{itemize}

\noindent 
The uncertainty associated with each ODE coefficient is effectively quantified through GP interpolation. While it is possible to utilize the predictive GP mean $m^{(*)}_{j,k}$ for each coefficient to obtain a mean prediction for FOM, an alternative approach is to use random samples. 
This approach yields multiple sets of ODEs for a single test point $\pmb{\mu}^{(*)}$, resulting in multiple corresponding latent space dynamics, each with varying likelihood. For instance, considering linear and constant candidate terms, we can generate $N_s$ sets of ODEs: 
\begin{equation}
    \begin{cases}
    \displaystyle\frac{dz^{(*)}_{:, 1}}{dt}=c_{1,1}^{(d)}z^{(*)}_{:, 1}+\cdots+c_{1,N_z}^{(d)}z^{(*)}_{:, N_z}+c_{1,N_l}^{(d)}\\[5pt]
    \,\,\,\,\,\vdots\\[5pt]
    \displaystyle\frac{dz^{(*)}_{:, N_z}}{dt}=c_{N_z,1}^{(d)}z^{(*)}_{:, 1}+\cdots+c_{N_z,N_z}^{(d)}z^{(*)}_{:, N_z}+c_{N_z,N_l}^{(d)}\\[5pt]
    \end{cases}
\end{equation}
with:
\begin{equation}
\label{sample_pred}
c^{(d)}_{j,k}\sim\mathcal{N}(m^{(*)}_{j,k},s^{(*)2}_{j,k})
\end{equation}

Next, we proceed to solve the corresponding latent space dynamics $\Tilde{\mathbf{Z}}^{(*,d)}$ for each sample $d\in[\![1,N_s]\!]$ (in the following sections, we use $N_s=20$). By making forward passes through the decoder network, we can estimate the uncertainty over the Full Order Model (FOM) prediction:
\begin{equation}
    \mathbb{E}[\tilde{\mathbf{u}}_n^{(*)}]=\frac{1}{N_s}\sum_{d=1}^{N_s}\phi_\text{dec}(\Tilde{\mathbf{z}}^{(*,d)}_n\,|\,\pmb{\theta}_\text{dec})\hspace{0.3in}n\in[\![0,N_t]\!]
\end{equation}
\begin{equation}
\label{pred_var}
    \mathbb{V}[\tilde{\mathbf{u}}_n^{(*)}]=\frac{1}{N_s}\sum_{d=1}^{N_s}(\phi_\text{dec}(\Tilde{\mathbf{z}}^{(*,d)}_n\,|\,\pmb{\theta}_\text{dec})-\mathbb{E}[\tilde{\mathbf{u}}_n^{(*)}])^2\hspace{0.3in}n\in[\![0,N_t]\!]
\end{equation}
The expected solution and variance of the ROM are denoted as $\mathbb{E}[\Tilde{\mathbf{U}}^{(*)}]=[\mathbb{E}[\Tilde{\mathbf{u}}^{(*)}_0],\dots,\mathbb{E}[\Tilde{\mathbf{u}}^{(*)}_{N_t}]]$ and $\mathbb{V}[\Tilde{\mathbf{U}}^{(*)}]=[\mathbb{V}[\Tilde{\mathbf{u}}^{(*)}_0],\dots,\mathbb{V}[\Tilde{\mathbf{u}}^{(*)}_{N_t}]]$, respectively.

\subsection{Variance--based Greedy Sampling}
\label{greedy}
\noindent 
In most applications, the accuracy of machine learning models, including the one described in this paper, is expected to improve with the availability of additional training data \cite{bishop2007}. Hence, selecting an appropriate parameter value, denoted as $\pmb{\mu}^{(N_\mu+1)}$, for running a FOM simulation and acquiring more training data is crucial. One intuitive approach is to choose a value of $\pmb{\mu}^{(N_\mu+1)}$ that hinders the most the model's ability from providing satisfactory accuracy. To achieve this, it is necessary to accurately assess the model's performance for any given $\pmb{\mu}^{(*)}$. In the gLaSDI framework \cite{glasdi}, this evaluation is accomplished by examining the decoder predictions for values of $\pmb{\mu}$ sampled from a discretized grid of the parameter space, denoted as $\mathcal{D}^h\subseteq\mathcal{D}$. These predictions are then incorporated into the PDE residual. However, this approach requires explicit knowledge of the PDE and can be computationally expensive and cumbersome to implement. In GPLaSDI, we propose an alternative method based on the uncertainty associated with the decoder predictions. Specifically, if the variance, denoted as $\mathbb{V}[\Tilde{\mathbf{U}}^{(*)}]$, is large compared to other points in the parameter space, then the expected value, denoted as $\mathbb{E}[\Tilde{\mathbf{U}}^{(*)}]$, is more likely to be inaccurate. Consequently, we select $\pmb{\mu}^{(N_\mu+1)}$ such that:
\begin{equation}
\label{argvar}
    \pmb{\mu}^{(N_\mu+1)}=\arg\max_{\pmb{\mu}^{(*)}\in\mathcal{D}^h}\Big[\max_{(t,x)}\mathbb{V}[\Tilde{\mathbf{U}}^{(*)}]\Big]
\end{equation}
In this paper, we evaluate $\mathbb{V}[\Tilde{\mathbf{U}}^{(*)}]$ at each and every point $\pmb{\mu}^{(*)}$ of $\mathcal{D}^h$. An alternative approach, faster, but less accurate, is to use only a handful of random points within $\mathcal{D}^h$ \cite{glasdi}. As for the acquisition of new data, we adopt a consistent sampling rate (i.e., every fixed $N_{up}$ training iterations). The steps involved in GPLaSDI, as discussed in Section \ref{gplasdi}, are summarized in Algorithm \ref{alg:cap}.

\begin{algorithm}
\caption{Autoencoder Training with Variance--based Greedy Sampling}\label{alg:cap}
\begin{algorithmic}[1]
\Require $\mathbf{U}=[\mathbf{U}^{(1)},\dots,\mathbf{U}^{(N_\mu)}]$, $N_\mu$, $N_{epoch}$, $N_z$, $N_l$, $N_{up}$, $\alpha$, $\beta_1$, $\beta_2$, $\beta_3$, $\pmb{\Theta}(\cdot)$, $\mathcal{D}^h$, $\phi_\text{enc}$, $\phi_\text{dec}$
\State Initialize $\pmb{\theta}_\text{enc}$, $\pmb{\theta}_\text{dec}$, $\pmb{\Xi}$, $\pmb{\theta}_\text{gp}^{j,k}$ randomly, and $h=0$
\While{$h<N_{epoch}$}
    \State Compute $\mathbf{Z}=\phi_\text{enc}(\mathbf{U}\,|\,\pmb{\theta}_\text{enc})$ and $\hat{\mathbf{U}}=\phi_\text{dec}(\mathbf{Z}\,|\,\pmb{\theta}_\text{dec})$ (See Eq. \eqref{ae})
    \State Compute $\mathcal{L}(\pmb{\theta}_\text{enc}, \pmb{\theta}_\text{dec}, \pmb{\Xi})$ (See Eq. \eqref{loss})
    \State Update $\pmb{\theta}_\text{enc}$, $\pmb{\theta}_\text{dec}$, and $\pmb{\Xi}$ using Adam algorithm, $\alpha$ and $\nabla\mathcal{L}(\pmb{\theta}_\text{enc}, \pmb{\theta}_\text{dec}, \pmb{\Xi})$
    \If{$h\mod N_{up}\equiv 0$}
        \For{$(j,k)\in[\![1,N_z]\!]\times[\![1,N_l]\!]$}
            \State Build dataset $(X_{j,k},y_{j,k})=(\pmb{\mu}^{(i)},\{\xi_{j,k}^{(i)}\})_{i\in[\![1,N_\mu]\!]}$
            \State Find $\pmb{\theta}_\text{gp}^{j,k}=\arg\min \mathcal{L}_\text{GP}(\pmb{\theta}_\text{gp})$ (See Eq. \eqref{mll})
        \EndFor
        \For{$\pmb{\mu}^{(*)}\in\mathcal{D}^h$}
            \For{$(j,k)\in[\![1,N_z]\!]\times[\![1,N_l]\!]$} 
                \State Compute $\{m^{(*)}_{j,k},s^{(*)}_{j,k}\}$ (See Eq. \ref{gp_pred2})
                \For{$d\in[\![1,N_s]\!]$}
                    \State Sample $c^{(d)}_{j,k}\sim\mathcal{N}(m^{(*)}_{j,k},s^{(*)2}_{j,k})$ (See Eq. \eqref{sample_pred})
                \EndFor
            \EndFor
            \For{$d\in[\![1,N_s]\!]$}
                \State Build the system of ODEs using $\{c^{(d)}_{j,k}\}_{(j,k)\in[\![1,N_z]\!]\times[\![1,N_l]\!]}$
                \State Solve for $\Tilde{\mathbf{Z}}^{(*,d)}$ and evaluate $\Tilde{\mathbf{U}}^{(*,d)}=\phi_\text{dec}(\Tilde{\mathbf{Z}}^{(*,d)}\,|\,\pmb{\theta}_\text{dec})$
            \EndFor
            \State Compute $\displaystyle\max_{(t,x)}\mathbb{V}[\Tilde{\mathbf{U}}^{(*)}]$ (See Eq. \eqref{pred_var})
        \EndFor
        \State Find $\displaystyle\pmb{\mu}^{(N_\mu+1)}=\arg\max_{\mathbf{\mu}^{(*)}\in \mathcal{D}^h}\Big[\max_{(t,x)}\mathbb{V}[\Tilde{\mathbf{U}}^{(*)}]\Big]$ (See Eq. \eqref{argvar})
        \State Collect $\mathbf{U}^{(N_\mu+1)}$ by running the FOM
        \State Update $\mathbf{U}=[\mathbf{U}^{(1)},\dots,\mathbf{U}^{(N_\mu)},\mathbf{U}^{(N_\mu+1)}]$ and $N_\mu=N_\mu+1$
    \EndIf
    \State Update $h=h+1$
\EndWhile

\end{algorithmic}
\end{algorithm}

\FloatBarrier

\section{Application}
\label{example}

\noindent In the following sections, we demonstrate the effectiveness of our method on multiple examples. To quantify the accuracy of GPLaSDI, we use the maximum relative error, defined as:
\begin{equation}
    e(\tilde{\mathbf{U}}^{(*)},\mathbf{U}^{(*)})=\max_n\bigg(\frac{|\!|\tilde{\mathbf{u}}^{(*)}_n-\mathbf{u}^{(*)}_n|\!|_2}{|\!|\mathbf{u}^{(*)}_n|\!|_2}\bigg)
\end{equation}
where $\tilde{\mathbf{u}}^{(*)}_n$ and $\mathbf{u}^{(*)}_n$ are the decoder ROM prediction and the ground truth at each time step, respectively (with GPLaSDI, we take $\tilde{\mathbf{u}}^{(*)}_n\equiv\mathbb{E}[\tilde{\mathbf{u}}^{(*)}_n]$). Each example are trained and tested on a compute node of the Livermore Computing Lassen supercomputer at the Lawrence Livermore National Laboratory, using a NVIDIA V100 (Volta) 64Gb GDDR5 GPU. Our GPLaSDI model is implemented using the open--source libraries \texttt{PyTorch} \cite{NEURIPS2019_9015} (for the autoencoder component) and \texttt{sklearn} (for the GP component) \cite{pedregosa2011scikit}. All the numerical examples shown in this paper can be regenerated by our open source code, GPLaSDI, which is available at GitHub page, i.e., \url{https://github.com/LLNL/GPLaSDI}.

\subsection{1D Burgers Equation}
\label{burgers1}
\noindent We first consider the inviscid 1D Burgers Equation, which was initially introduced in \cite{lasdi} and further discussed in \cite{glasdi}:
\begin{equation}
    \begin{cases}
    \displaystyle\frac{\partial u}{\partial t}+u\frac{\partial u}{\partial x}=0\hspace{0.3in}(t,x)\in[0,1]\times[-3,3]\\
    \displaystyle u(t,x=3)=u(t,x=-3)
    \end{cases}
\end{equation}
The initial condition is parameterized by $\pmb{\mu}=\{a,w\}\in\mathcal{D}$, and the parameter space is defined as $\mathcal{D}=[0.7,0.9]\times[0.9,1.1]$:
\begin{equation}
    u(t=0,x)=a\exp\bigg(-\frac{x^2}{2w^2}\bigg)\hspace{0.3in}\pmb{\mu}=\{a,w\}\\
\end{equation}
The FOM solver utilizes an implicit backward Euler time integration scheme and a backward finite difference discretization in space. The spatial stepping is set to $\Delta x=6\cdot10^{-3}$ and the time stepping to $\Delta t=10^{-3}$. In this section and in section \ref{burgers2}, the discretization is based on the one used in \cite{lasdi, glasdi}, and is chosen to ensure stability at any point of the parameter space. For the purpose of greedy sampling, the parameter space is discretized into a square grid $\mathcal{D}^h$ with a stepping of $\Delta a=\Delta w=0.01$, resulting in a total of 441 grid points ($21$ values in each dimension). The initial training dataset consists of $N_\mu=4$ FOM simulations, corresponding to the parameters located at each corner of $\mathcal{D}^h$. Specifically, the parameter values are $\pmb{\mu}^{(1)}=\{0.7,0.9\}$, $\pmb{\mu}^{(2)}=\{0.9,0.9\}$, $\pmb{\mu}^{(3)}=\{0.7,1.1\}$ and $\pmb{\mu}^{(4)}=\{0.9,1.1\}$.

The encoder architecture follows a 1001--100--5 structure, comprising one hidden layer with $100$ hidden units and $N_z=5$ latent variables. The decoder has a symmetric configuration to the encoder. It employs a sigmoid activation function. Note that in this paper, we select the autoencoder architecture through a random search using only the initial four corner training datapoints, on trainings that are run for at most $N_{up}$ epochs (i.e. no sampling of additional FOM data). This is done to ensure that the training loss does not get stuck and that the learning rate is appropriate. The cost of running these search is cheap since there is few training points and no FOM sampling. Throughout each examples, as a rule of thumb, we have found that decreasing the width of each hidden layer by an order of magnitude compared to the previous layer generally yields satisfactory performance. It is possible that other architectures may lead to equivalent, better, or worse performance (e.g. convolutional layers, etc.). \\
To identify the latent space dynamics, the dictionary of possible terms $\pmb{\Theta}(\cdot)$ includes only linear terms and a constant, resulting in $N_l=6$ terms. The autoencoder is trained for $N_{epoch}=2.8\cdot10^4$ iterations, with a learning rate $\alpha=10^{-3}$. A new FOM data point is sampled every $N_{up}=2000$ iterations (resulting in adding 13 data points during training, for a total of 17 training points). For estimating the prediction variance, $20$ samples are used (i.e., $N_s=20$). The loss hyperparameters are set as $\beta_1=1$, $\beta_2=0.1$, and $\beta_3=10^{-6}$. The hyperparameters employed in this example are based on \cite{glasdi}. Additional details on the effects of hyperparameter selection can be found in \cite{glasdi, lasdi}.

\begin{figure}
\vspace{-4.5cm}
\hspace{-6cm}
    \includegraphics[width=2\textwidth]{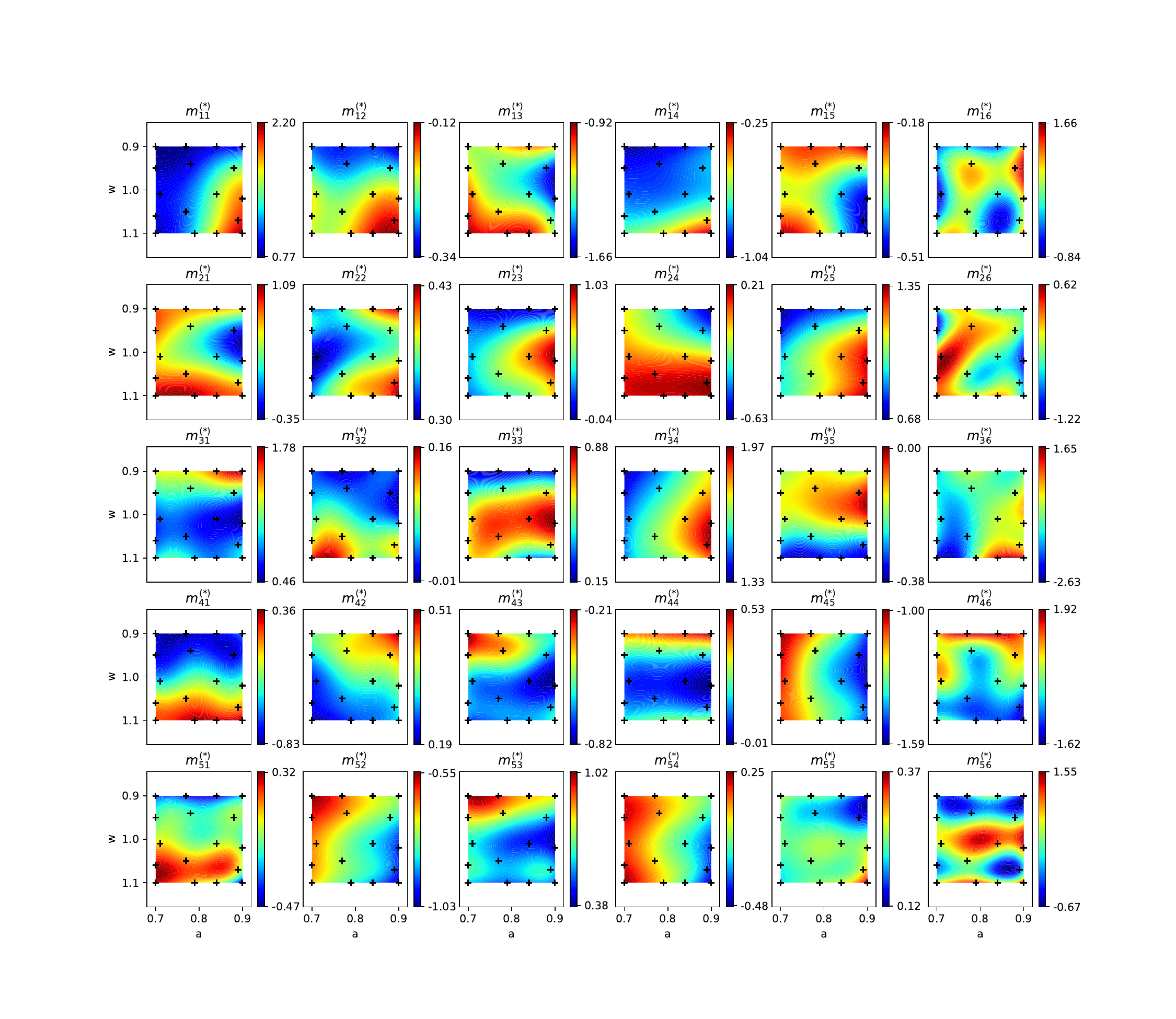}
    \caption{1D Burgers Equation -- Predictive mean of each ODE coefficients ($m_{j,k}^{(*)}$) given $\pmb{\mu}^{(*)}$ at the end of the training. The black marks represent each sampled data point.}
    \label{burgers1_gp_mean}
\end{figure}

\begin{figure}
\vspace{-4.5cm}
\hspace{-6cm}
    \includegraphics[width=2\textwidth]{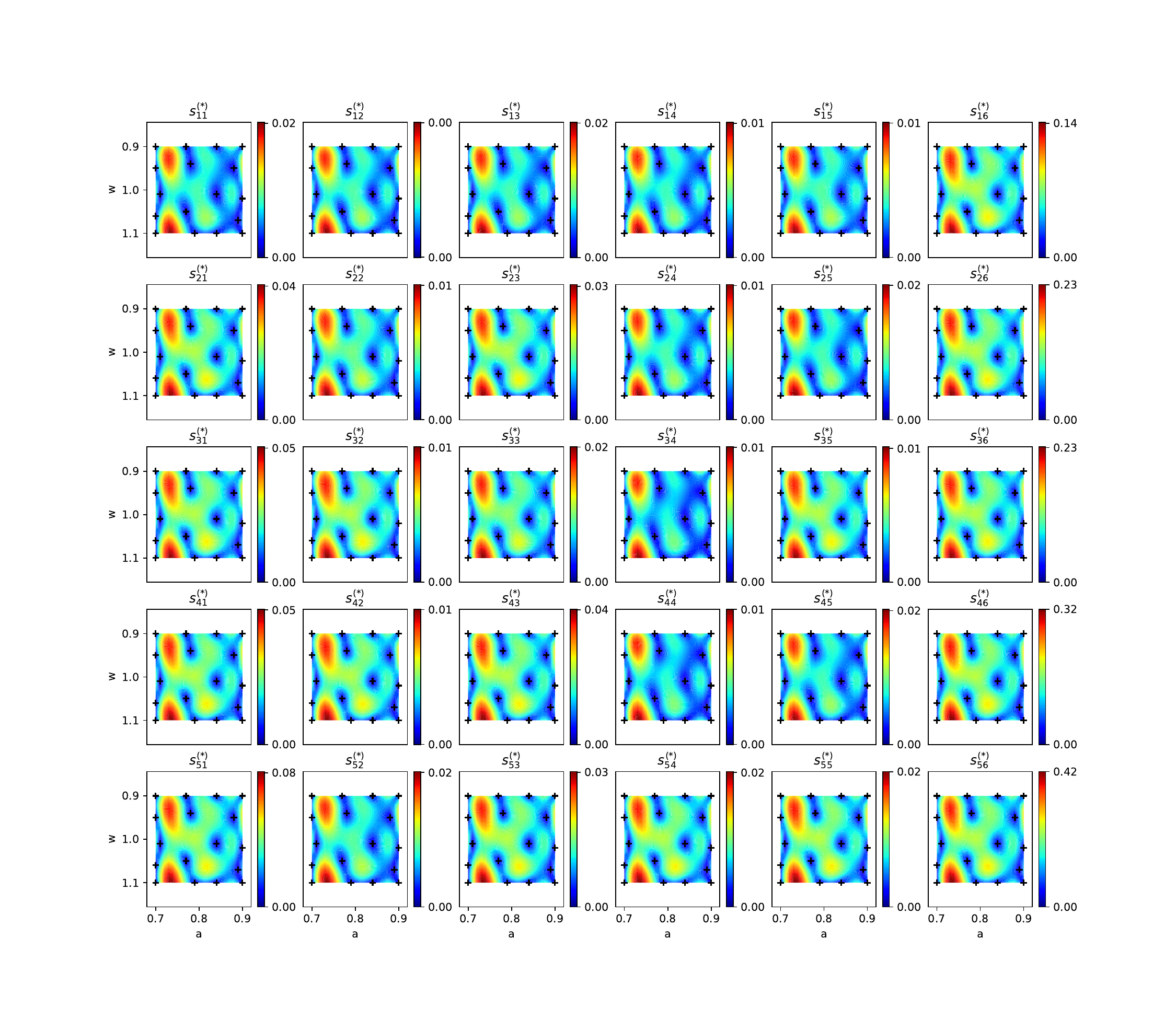}
    \caption{1D Burgers Equation -- Predictive standard deviation of each ODE coefficients ($s_{j,k}^{(*)}$) given $\pmb{\mu}^{(*)}$. The heatmaps are similar for each coefficients, which is expected because here the input point locations within $\mathcal{D}^h$ are all the same. Notice that the uncertainty is higher in regions with no training data points, as one might intuitively expect.}
    \label{burgers1_gp_std}
\end{figure}

\begin{figure}
\vspace{-3.5cm}
\hspace{-1.8cm}
    \includegraphics[width=1.3\textwidth]{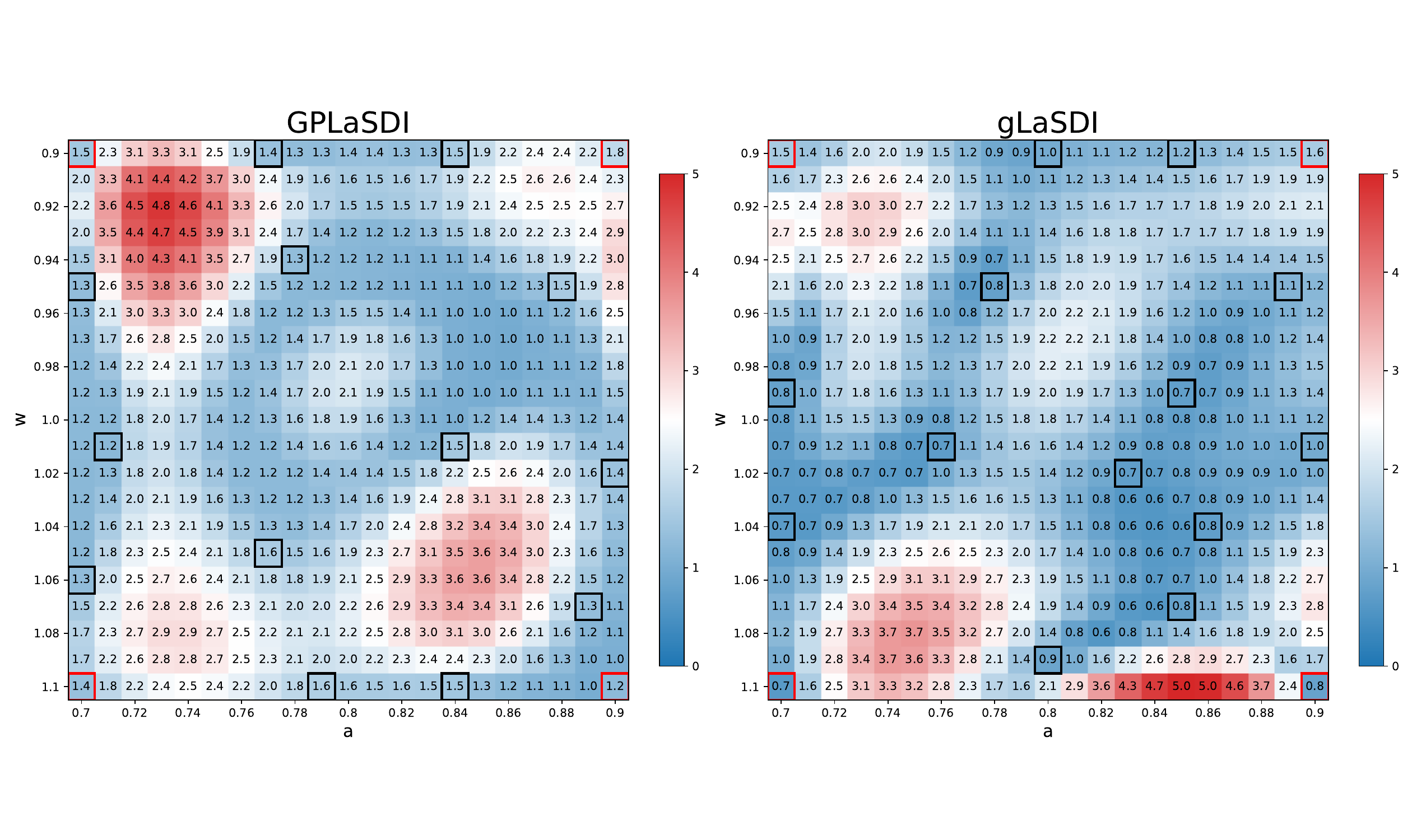}
    \caption{1D Burgers Equation -- Maximum relative error ($\%$) using GPLaSDI (left) and gLaSDI (right). The values in a red square correspond to the original FOM data at the beginning of the training (located at the four corners). The values in a black square correspond to parameters and FOM runs that were sampled during training. 
    }
    \label{burgers1_max_error_mean}
\end{figure}

\begin{figure}
\hspace{1.5cm}
    \includegraphics[width=0.8\textwidth]{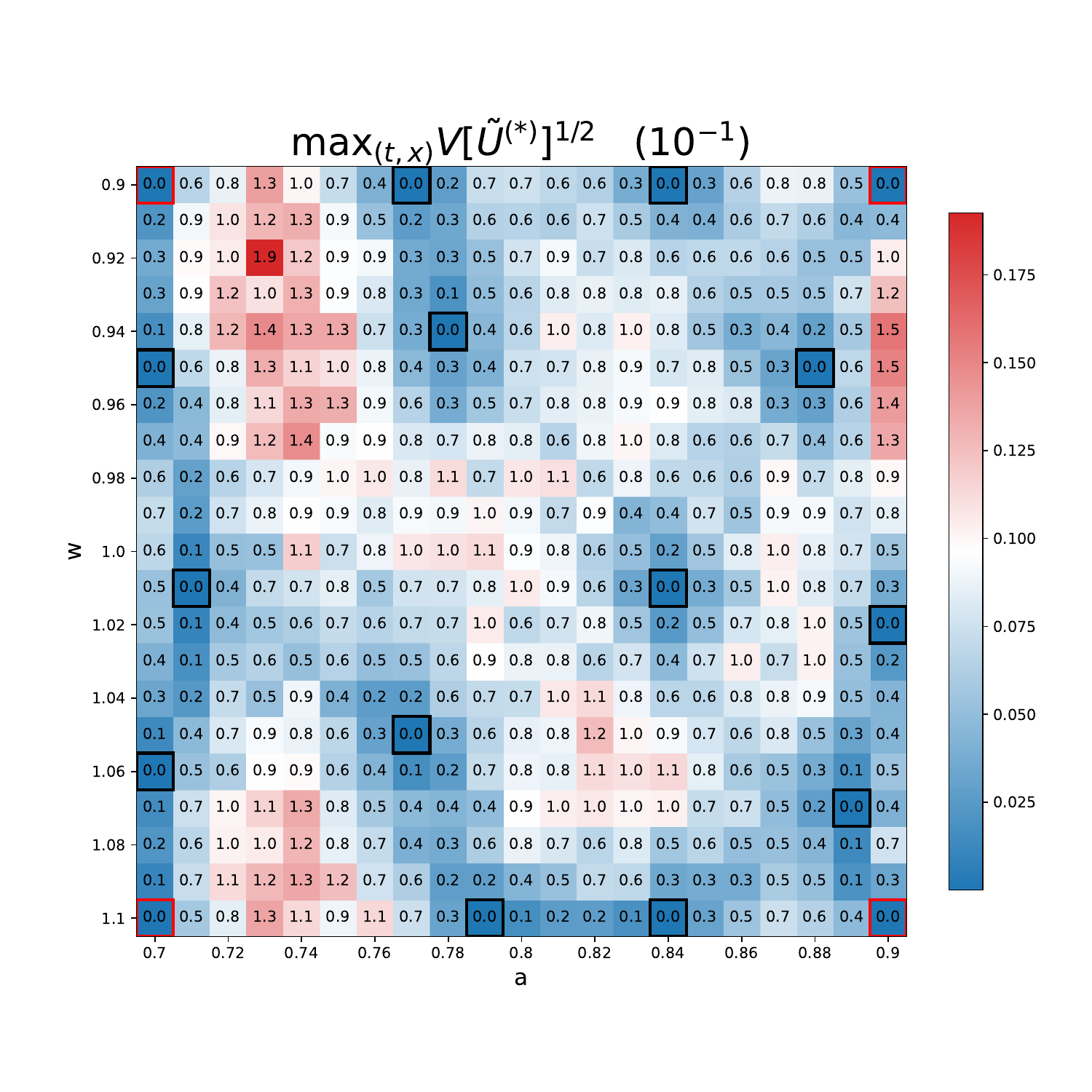}
    \caption{1D Burgers Equation -- Maximum predictive standard deviation for GPLaSDI. The numbers inside each box are in scientific notations, scaled by the factor of 10 specified in the title. For example, the maximum value across the figure is $1.9\cdot10^{-1}$.}
    \label{burgers1_max_std}
\end{figure}

\begin{figure}[!h]
\hspace{-1cm}
    \includegraphics[width=1.2\textwidth]{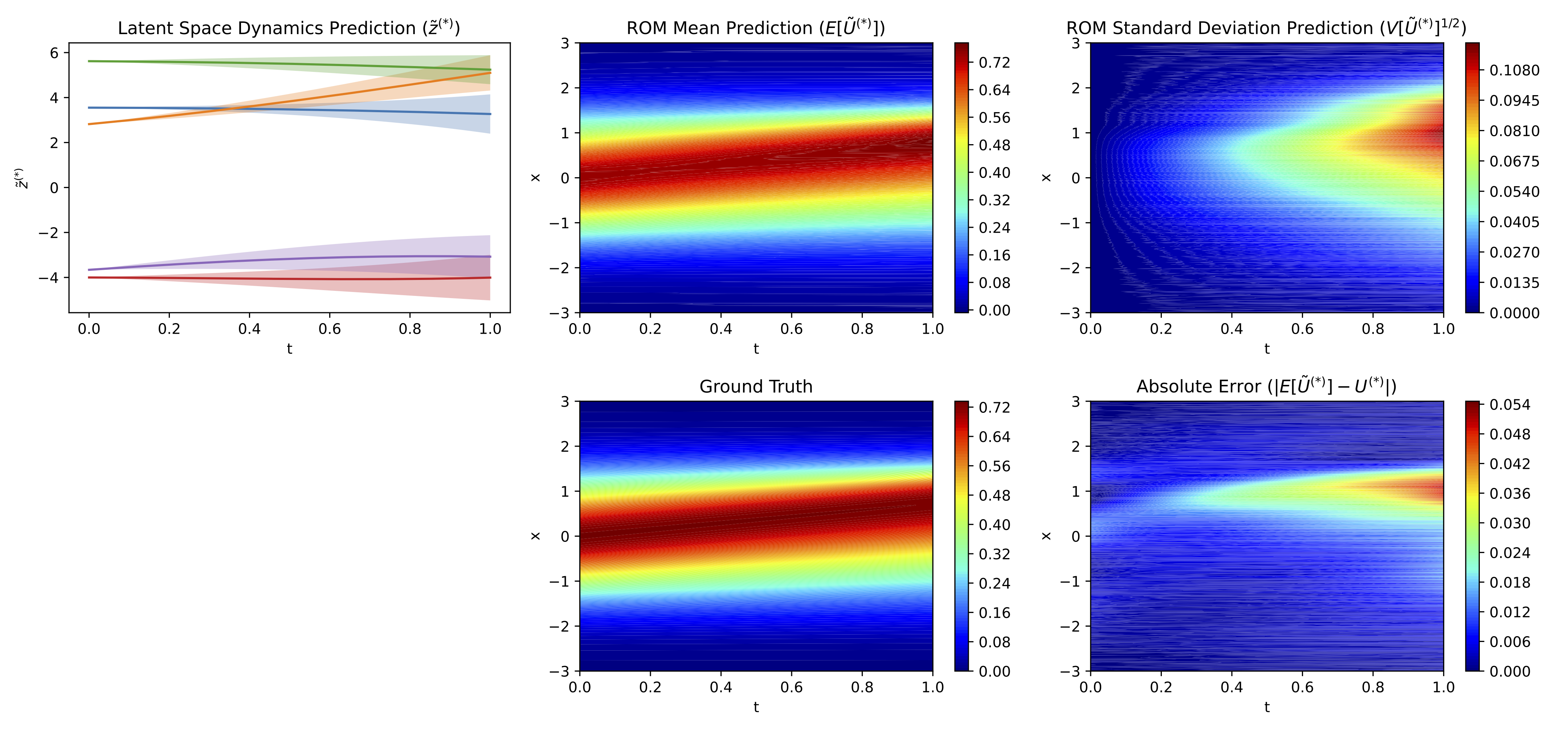}
    \caption{1D Burgers Equation -- Predictions for $\pmb{\mu}^{(*)}=\{0.73,0.92\}$. This plot shows the predicted latent space dynamics $\mathbb{E}[\mathbf{\tilde{Z}}^{(*)}]$ with a $95\%$ confidence interval, the ROM mean prediction $\mathbb{E}[\mathbf{\tilde{U}}^{(*)}]$ (decoder output) and standard deviation, the ground truth, and the absolute error. 
    }
    \label{burgers1_prediction}
\end{figure}

For baseline comparison, a gLaSDI model \cite{glasdi} is also trained using identical hyperparameter settings, autoencoder architecture, and GP interpolation of the latent space dynamics. The key difference lies in the data sampling strategy: gLaSDI employs PDE residual error--based sampling, while GPLaSDI utilizes uncertainty--based sampling.
\\\\
\noindent The system of governing ODEs for the latent space dynamics consists of 30 coefficients, each of which is interpolated by a GP. Figure \ref{burgers1_gp_mean} illustrates the predictive mean of each GP, while Figure \ref{burgers1_gp_std} displays the corresponding standard deviation. Figure \ref{burgers1_max_error_mean} showcases the maximum relative error (in percentage) for each test point in $\mathcal{D}^h$, obtained using GPLaSDI and gLaSDI \cite{glasdi}. Remarkably, our GPLaSDI framework achieves outstanding performance, with the worst maximum relative error being below $5\%$, and the majority of predictions exhibiting less than $3.5\%$ error. Moreover, for this particular example, GPLaSDI slightly outperforms gLaSDI, where the worst maximum relative error reaches $5\%$.
\\\\
\noindent In Figure \ref{burgers1_max_std}, the maximum standard deviation $\max_{(t,x)}\mathbb{V}[\tilde{\mathbf{U}}^{(*)}]^{1/2}$ is depicted, representing the uncertainty in the decoder ROM predictions. As expected, the standard deviation patterns closely match the predictive standard deviation of each GP (Figure \ref{burgers1_gp_std}). There is also a significant correlation between high standard deviation and high maximum relative error (Figure \ref{burgers1_max_error_mean}). This correlation is valuable as it indicates that the uncertainty in the ROM model can be reliably quantified. This observation may also explain why GPLaSDI performs similarly to gLaSDI. A ROM that incorporates knowledge of the underlying physics is generally expected to outperform purely data-driven models due to additional insights. However in this case, GPLaSDI, despite being agnostic to the PDE (unlike gLaSDI), effectively captures the correlation between prediction uncertainty and ROM error. It therefore serves as a robust surrogate for the PDE residual error.\\
Figure \ref{burgers1_prediction} illustrates the model prediction for parameter $\pmb{\mu}^{(*)}=\{0.73,0.92\}$, corresponding to the case with the largest relative error. Despite the larger error, the model predictions remain reasonable, and there is a clear correlation between the predictive standard deviation and the absolute error compared to the ground truth. The absolute error consistently falls within one standard deviation.
\\\\
\noindent During $50$ test runs, the FOM wall clock run-time averages at $1.31$ seconds using a single core. On the other hand, the ROM model requires an average of $6.36\cdot10^{-3}$ seconds, resulting in an average speed-up of $206\times$. It is important to note that in this case, the system of ODEs is solved using the GP predictive means ($m^{(*)}_{j,k}$), requiring only one integration of the system of ODEs and one forward pass through the decoder. This approach does not allow for predictive standard deviation estimations since multiple sets of ODEs would need to be solved, and an equivalent number of forward passes would be required. Therefore, if we aim to make a ROM prediction with uncertainty estimation using, for example, 10 samples, the speed-up would reduce to roughly $20\times$. Note that running the ROM predictions for multiple samples could also be done in an embarrassingly parallel way, which would limit the deterioration of speed-up performances.
\\\\
\noindent As in any neural-network-based algorithm, precisely understanding the source of errors can be challenging. There exist however several ways of pinpointing and mitigating possible errors in GPLaSDI:
\begin{itemize}
    \item Compute the maximum relative error for points belonging to the training set (i.e. reproductive case), where ground truth data is available. Note that the training data comes from deterministic numerical solvers with unchanged simulation settings, and can thus be considered as noiseless data (as opposed to experimental data for instance). Therefore, overfitting of the autoencoder is not a primary concern and is not expected to be a major source of error. On the other hand, underfitting and/or using an autoencoder that does not have an architecture capable of capturing all the complexity in the physics is always a risk. This can be easily assessed by looking at the autoencoder prediction error on the training data (e.g. mean-squared-error between $\mathbf{U}$ and $\mathbf{\hat{U}}$).
    \item Comparing $\mathbf{\hat{Z}}$ and $\mathbf{\tilde{Z}}$ through preliminary visual inspection and error metrics such as mean-squared-error. If the error between $\mathbf{U}$ and $\mathbf{\hat{U}}$ is low, but the error between $\mathbf{\hat{Z}}$ and $\mathbf{\tilde{Z}}$ is high, this would indicate that the autoencoder is well trained, but that the model fails to reproduce the dynamics of $\mathbf{\hat{Z}}$, and more emphasis needs to be put on the SINDy loss (i.e. increase $\beta_2$). 
    \item In this paper, to demonstrate the performance of GPLaSDI, we have generated FOM testing data for every single point in $\mathcal{D}^h$ (to compute the maximum relative error). In practice, doing so may be very expensive and would most likely defeat the purpose of a ROM. It may be possible however to generate only a handful of test FOM datapoints at random locations of the parameter space. Another way of estimating the error could be to feed the ROM predictions into the PDE residual and compute the residual error, if available. This would however defeat the purpose of a non-intrusive ROM, and in such case, using an intrusive approach such as \textit{gLaSDI} \cite{glasdi} may be desirable. Note that a hybrid algorithm, using both the PDE residual (for physics information) and GPs to interpolate the sets of ODEs (for uncertainty quantification) is perfectly conceivable, but left to future work.
\end{itemize}

\FloatBarrier

\FloatBarrier

\FloatBarrier
\subsection{2D Burgers Equation}
\label{burgers2}
\begin{figure}[!h]
\vspace{-3.cm}
\hspace{-1.5cm}
    \includegraphics[width=1.3\textwidth]{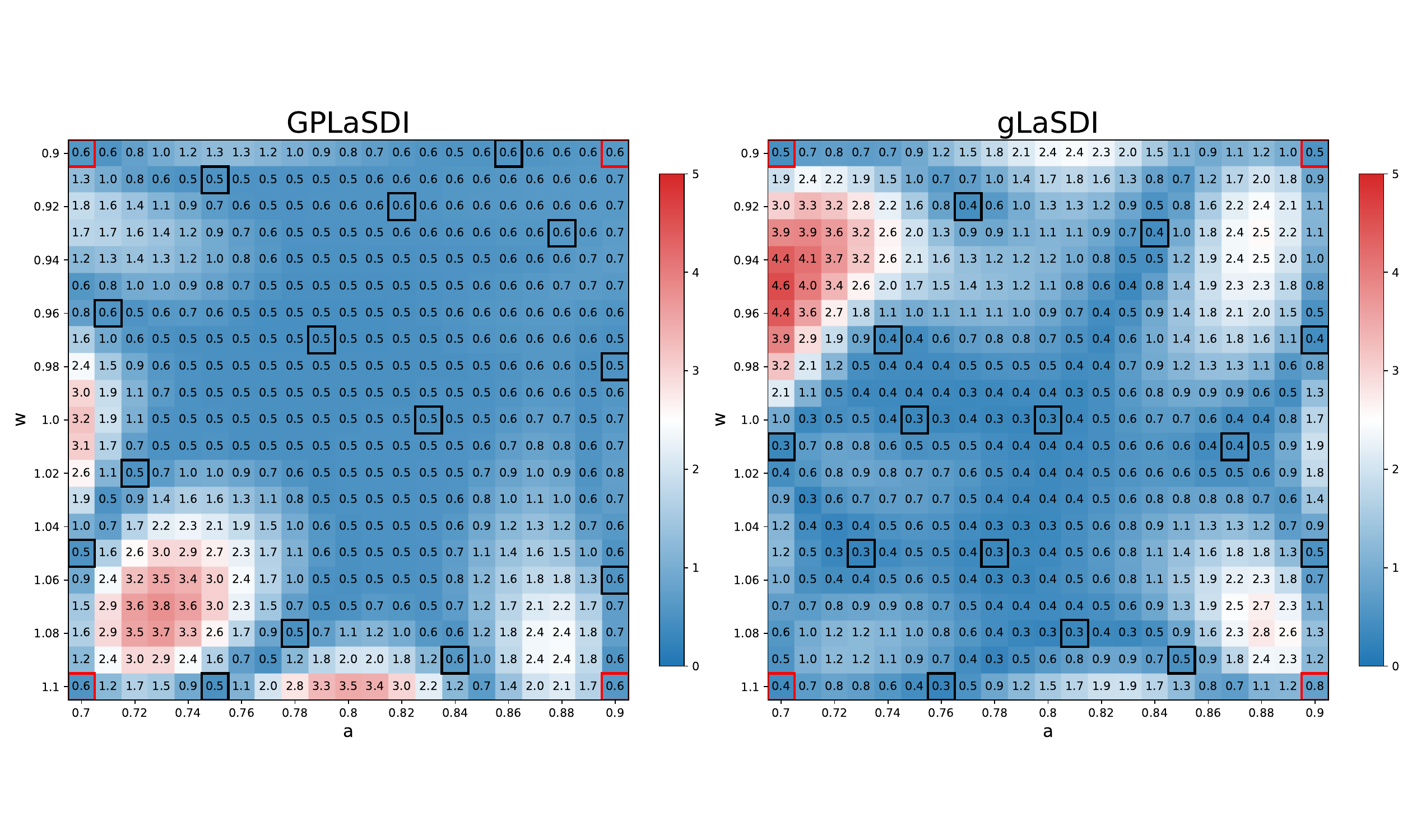}
    \caption{2D Burgers Equation -- Maximum relative error ($\%$) using GPLaSDI (left) and gLaSDI (right)).}
    \label{burgers2_max_error_mean}
\end{figure}

\noindent We now consider the 2D Burgers equation with a viscous term, as introduced in \cite{lasdi} and \cite{glasdi}:
\begin{equation}
\begin{cases}
    \displaystyle\frac{\partial \mathbf{u}}{\partial t}+\mathbf{u}\cdot\nabla\mathbf{u}=\frac{1}{Re}\Delta\mathbf{u}\hspace{0.3in}(t,x,y)\in[0,1]\times[-3,3]\times[-3,3]\\
    \displaystyle \mathbf{u}(t,x=\pm3,y=\pm3)=\mathbf{0}
\end{cases}
\end{equation}
The initial condition is analogous to the 1D case:
\begin{equation}
    u(t=0,x,y)=v(t=0,x,y)=a\exp\bigg(-\frac{x^2+y^2}{2w^2}\bigg)\hspace{0.3in}
    \begin{cases}
    \pmb{\mu}=\{a,w\}\\
    \mathbf{u}=\{u,v\}
    \end{cases}
\end{equation}
\noindent We consider a Reynolds number of $Re=10^5$. The FOM solver employs an implicit backward Euler time integration scheme, a backward finite difference for the nonlinear term, and a central difference discretization for the diffusion term. The spatial resolution is set to $\Delta x=\Delta y=0.1$, and the time stepping is set to $\Delta t=5\cdot10^{-3}$. These settings remain consistent with the 1D case. However, in this scenario, we employ a neural network architecture of 7200--100--20--20--20-5 for the encoder, and a symmetric architecture for the decoder. The activation function used is softplus. The autoencoder is trained for $N_{epoch}=7.5\cdot10^5$ iterations, with a sampling rate of $N_{up}=5\cdot10^4$ (resulting in adding 14 data points during training, for a total of 18 training points).
\\\\
\noindent Figure \ref{burgers2_max_error_mean} depicts the maximum relative error for each point in the parameter space, comparing the performance of GPLaSDI and gLaSDI. GPLaSDI achieves consistently low maximum relative errors across most of the parameter space, typically ranging from $0.5\%$ to $1\%$, with a maximum error not exceeding $3.8\%$. On the other hand, gLaSDI exhibits a larger maximum relative error, reaching up to $4.6\%$.

\begin{figure}[!h]
\hspace{1.5cm}
    \includegraphics[width=0.8\textwidth]{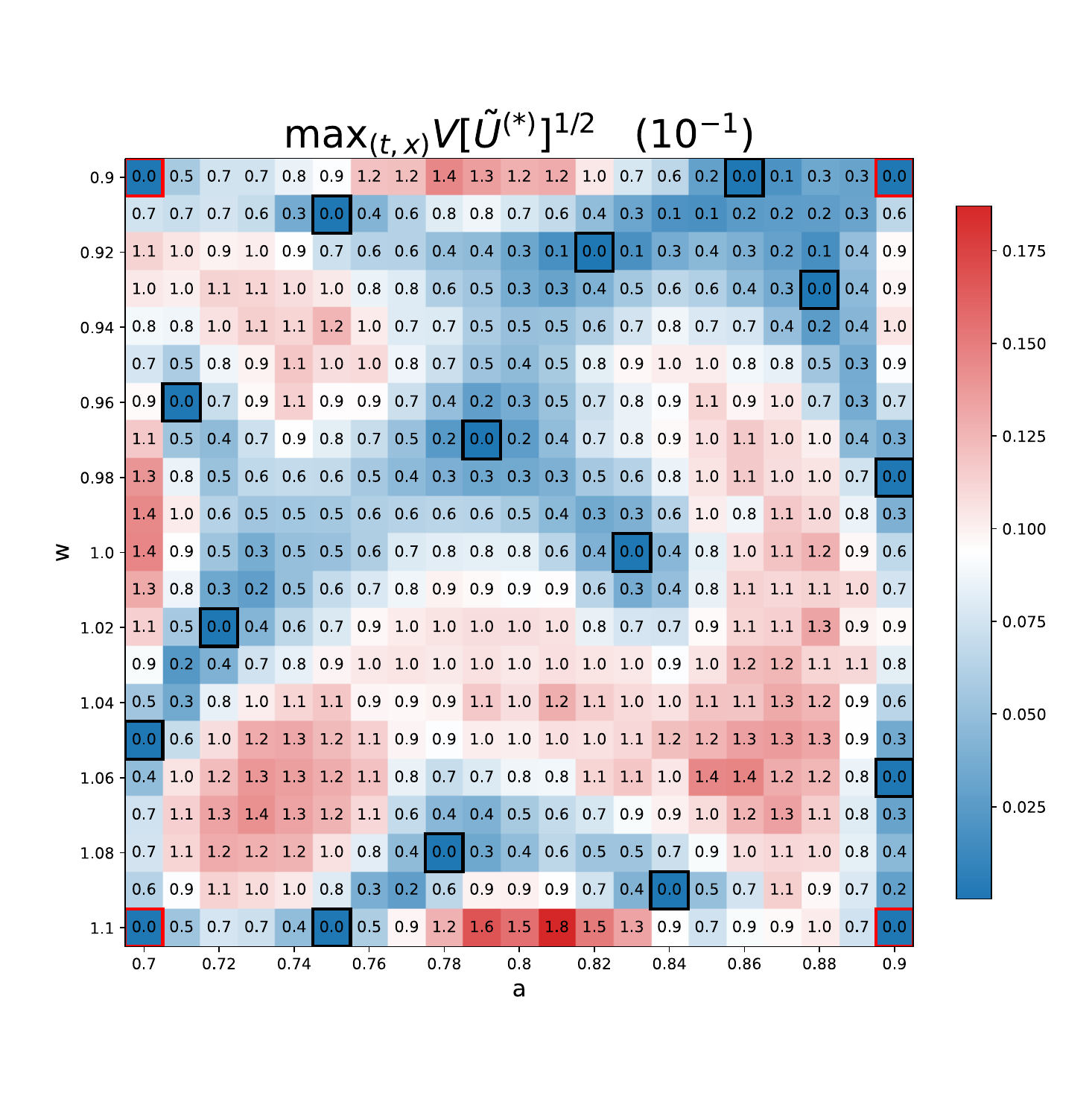}
    \caption{2D Burgers Equation -- Maximum predictive standard deviation for GPLaSDI.}
    \label{burgers2_max_std}
\end{figure}

In Figure \ref{burgers2_max_std}, we observe the maximum standard deviation, which, similar to the 1D case, exhibits a clear correlation with the maximum relative error.

\begin{figure}[!h]
\vspace{-1cm}
\hspace{-2.8cm}
        \includegraphics[width=1.5\textwidth]{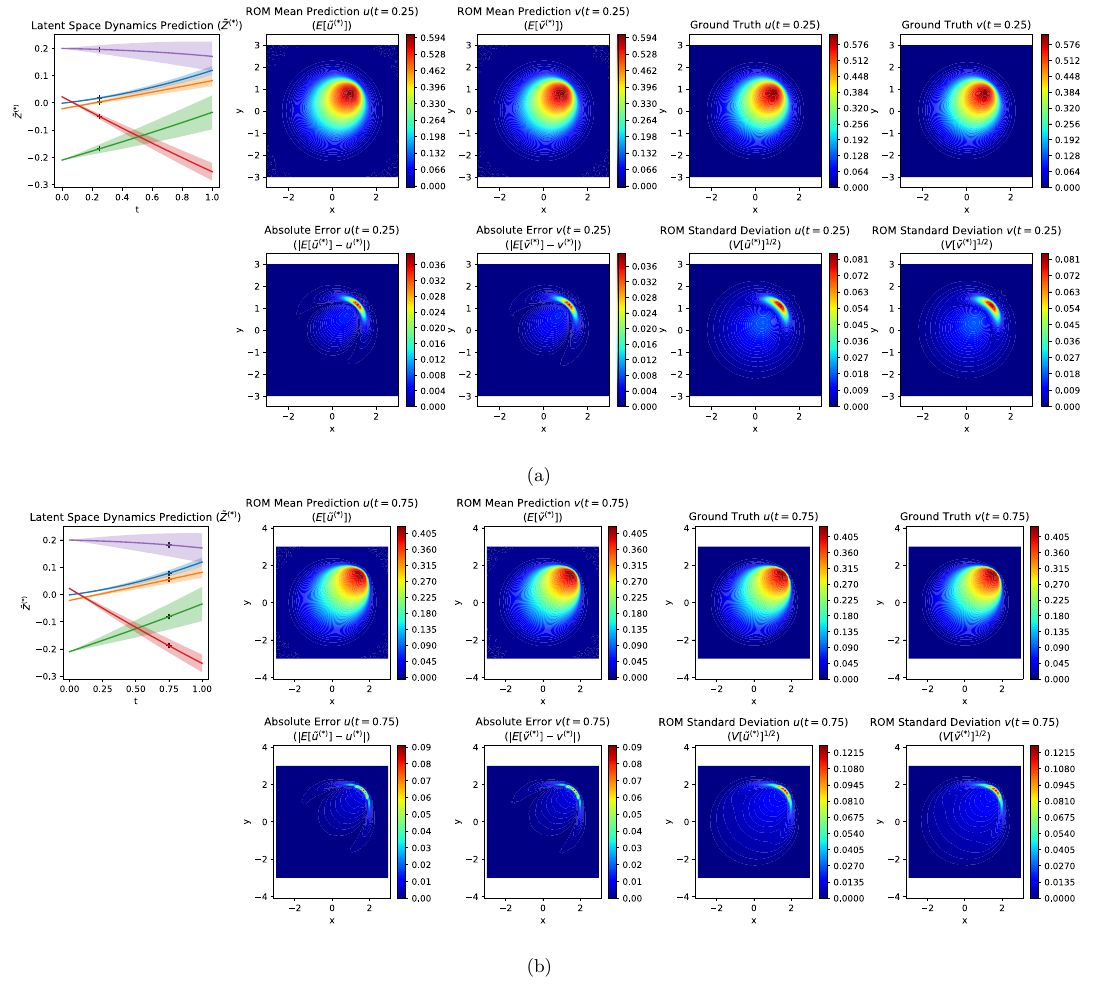}
    \caption{2D Burgers Equation -- Predictions for $\pmb{\mu}^{(*)}=\{0.73,1.07\}$ at $t=0.25$ (a) and $t=0.75$ (b). These plot show the predicted latent space dynamics $\mathbb{E}[\mathbf{\tilde{Z}}^{(*)}]$ with a $95\%$ confidence interval, the ROM mean prediction for $\mathbf{u}=\{u,v\}\equiv\{\mathbb{E}[\tilde{u}^{(*)}], \mathbb{E}[\tilde{v}^{(*)}]\}$ (decoder output) and standard deviation, the ground truth, and the absolute error. 
    }
    \label{burgers2_prediction}
\end{figure}

Figure \ref{burgers2_prediction} presents the dynamics of the latent space, the predicted and ground truth $\mathbf{u}$ fields, the absolute error, and the fields' predictive standard deviations for the least favorable case ($\pmb{\mu}^{(*)}=\{0.73,1.07\}$) at time points $t=0.25$ and $t=0.75$. The error predominantly concentrates along the shock front, where the discontinuity forms. However, the error remains well within the predictive standard deviation, affirming the capability of GPLaSDI to provide meaningful confidence intervals.

During $50$ test runs, the FOM wall clock run--time averaged $63.5$ seconds using a single core. In contrast, the ROM model achieved an average runtime of blue$9.13\cdot10^{-3}$ seconds, resulting in an average speed-up of $6949\times$. Similarly to the 1D case, this speed-up is attained solely using the mean prediction. To incorporate uncertainty prediction, additional ODE integrations and decoder forward passes would be required, thereby slightly diminishing the speed-up gains.

\FloatBarrier

\subsection{Two--Stream Plasma Instability}
\label{vlasov}
\noindent In this section, we consider the simplified 1D--1V Vlasov--Poisson equation:
\begin{equation}
\label{vlasov_eqn}
\begin{cases}
    \displaystyle\frac{\partial f}{\partial t}
    +\frac{\partial }{\partial x} \left(v f\right) 
    + \frac{\partial}{\partial v}\left(\frac{d \phi}{dx} f\right) 
    = 0,
    \hspace{0.3in}
    (t,x,v)\in[0,5]\times[0,2\pi]\times[-7,7],\\[5pt]
    \displaystyle\frac{d^2\phi}{dx^2} = \int_v f dv,
\end{cases}
\end{equation}
where we consider a plasma distribution function denoted as $f \equiv f\left(x,v\right)$, which depends on the variables $x$ (physical coordinate) and $v$ (velocity coordinate). The equation also involves the electrostatic potential $\phi$. This simplified model governs 1D collisionless electrostatic plasma dynamics and is representative of complex models for plasma behavior in various applications, including proposed fusion power plant designs. It is important to note that kinetic modeling of plasmas leads to high-dimensional PDEs; although Equation \eqref{vlasov_eqn} describes one-dimensional dynamics, it is a two-dimensional PDE.

In this example, we focus on solving the two-stream instability problem that is a canonical problem used to simulate the excitation of a plasma wave from counterstreaming plasmas. The initial solution is:
\begin{equation}
    f(t=0,x,v)=\frac{8}{\sqrt{2\pi T}}\bigg[1+\frac{1}{10}\cos(k\pi x)\bigg]\bigg[\exp\bigg(-\frac{(v-2)^2}{2T}\bigg)+\exp\bigg(-\frac{(v+2)^2}{2T}\bigg)\bigg],
\end{equation}

where $T$ represents the plasma temperature. The parameters involved are denoted as $\pmb{\mu}=\{T,k\}$, where $T$ ranges from $0.9$ to $1.1$, and $k$ ranges from $1.0$ to $1.2$. We discretize the parameter space over a $21\times21$ grid $\mathcal{D}^h$, with a step size of $\Delta T=\Delta k=0.01$. The training process is initialized with $N_\mu=4$ training data points located at the four corners of the parameter space. The FOM data is sampled using \texttt{HyPar} \cite{HyPar}, a conservative finite difference PDE code. It utilizes the fifth order WENO discretization~\cite{jiangshu} in space ($\Delta x=2\pi/128$, $\Delta v=7/128$) and the classical four--stage, fourth--order Runge--Kutta time integration scheme ($\Delta t=5\cdot10^{-3}$). In this section and in section \ref{2D Rising Thermal Bubble}, the discretization is based on the showcasing examples of \texttt{HyPar} \cite{HyPar}, and is chosen to ensure stability at any point of the parameter space.

For the neural network architecture, we use a 16384--1000--200--50--50--50--5 configuration for the encoder (and a symmetric architecture for the decoder). The activation function employed is softplus. The latent space consists of $N_z=5$ variables, and only linear terms are considered for the SINDy library. The loss hyperparameters are set as $\beta_1=1$, $\beta_2=0.1$, and $\beta_3=10^{-5}$. To estimate the prediction variance, we utilize $N_s=20$ samples. The training process involves $N_{epoch}=6.5\cdot10^5$ epochs with a learning rate of $\alpha=10^{-5}$ and $N_{up}=5\cdot10^4$ updates (resulting in adding 12 data points during training, for a total of 16 training points).

To provide a baseline for comparison, we evaluate the performance of GPLaSDI against an autoencoder trained with the same settings and hyperparameters on a uniform parameter grid. The uniform grid consists of $16$ data points arranged in a $4\times4$ grid. In the baseline model, the interpolations of the latent space ODE coefficients are performed using GPs. Similar to GPLaSDI, the training process for the baseline model also incorporates active learning. It begins with the initial four corner points, and at every $N_{up}=5\cdot10^4$ iterations, a new point is randomly selected from the uniform grid.

\begin{figure}[!h]
\hspace{-1.5cm}
    \includegraphics[width=1.3\textwidth]{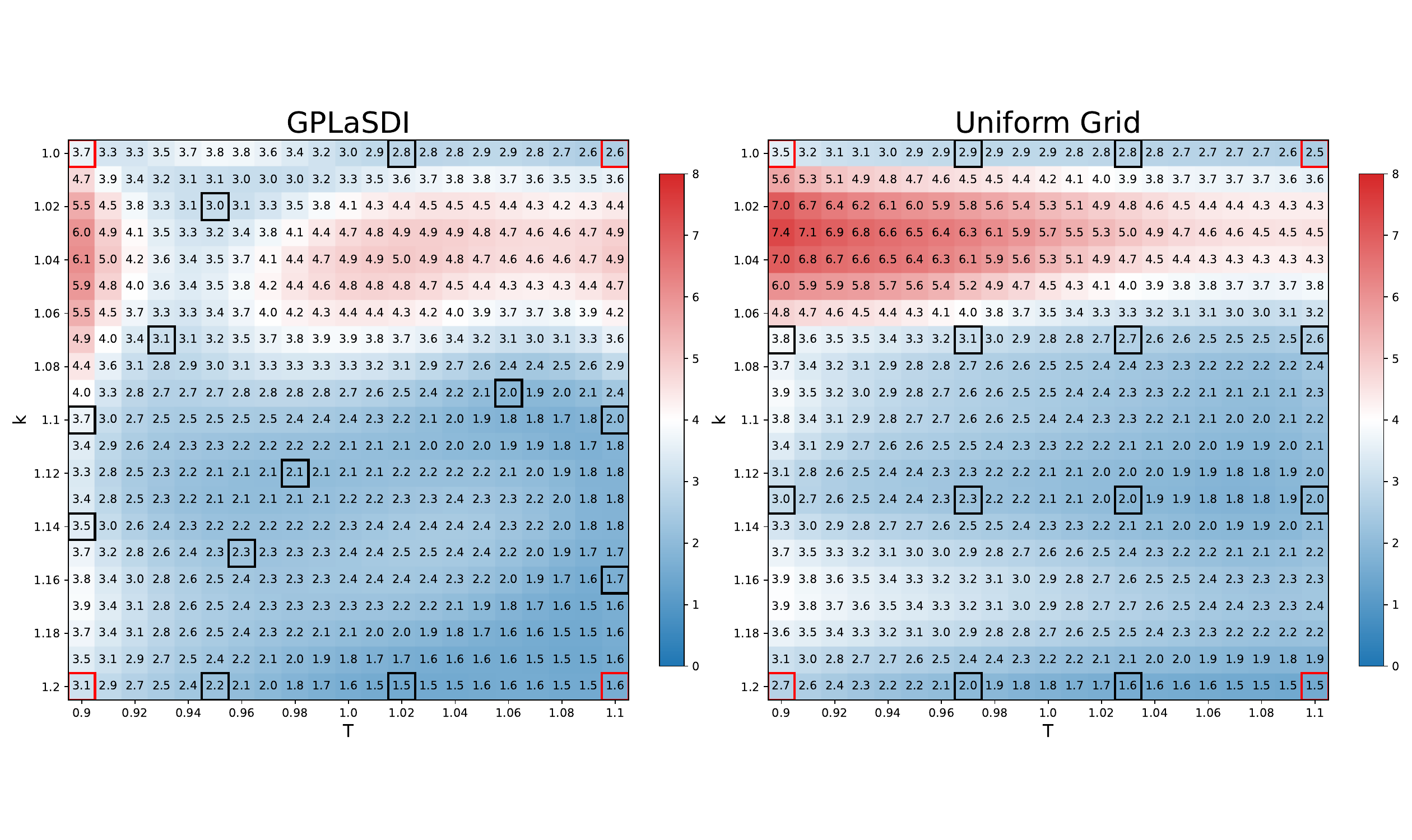}
    \caption{1D1V Vlasov Equation -- Maximum relative error ($\%$) using GPLaSDI and a uniform training grid (non--greedy).}
    \label{vlasov_max_error_mean}
\end{figure}

Figure \ref{vlasov_max_error_mean} presents the maximum relative error for each point in the parameter space obtained using GPLaSDI and the baseline model. With GPLaSDI, the worst maximum relative error is $6.1\%$, and in most regions of the parameter space, the error remains within the range of $1.5-3.5\%$. The highest errors are concentrated towards smaller values of $k$ (typically $k<1.07$). Compared to uniform sampling, GPLaSDI outperforms the baseline model, which achieves a maximum relative error of $7.4\%$.

\begin{figure}[!h]
\hspace{1cm}
    \includegraphics[width=0.8\textwidth]{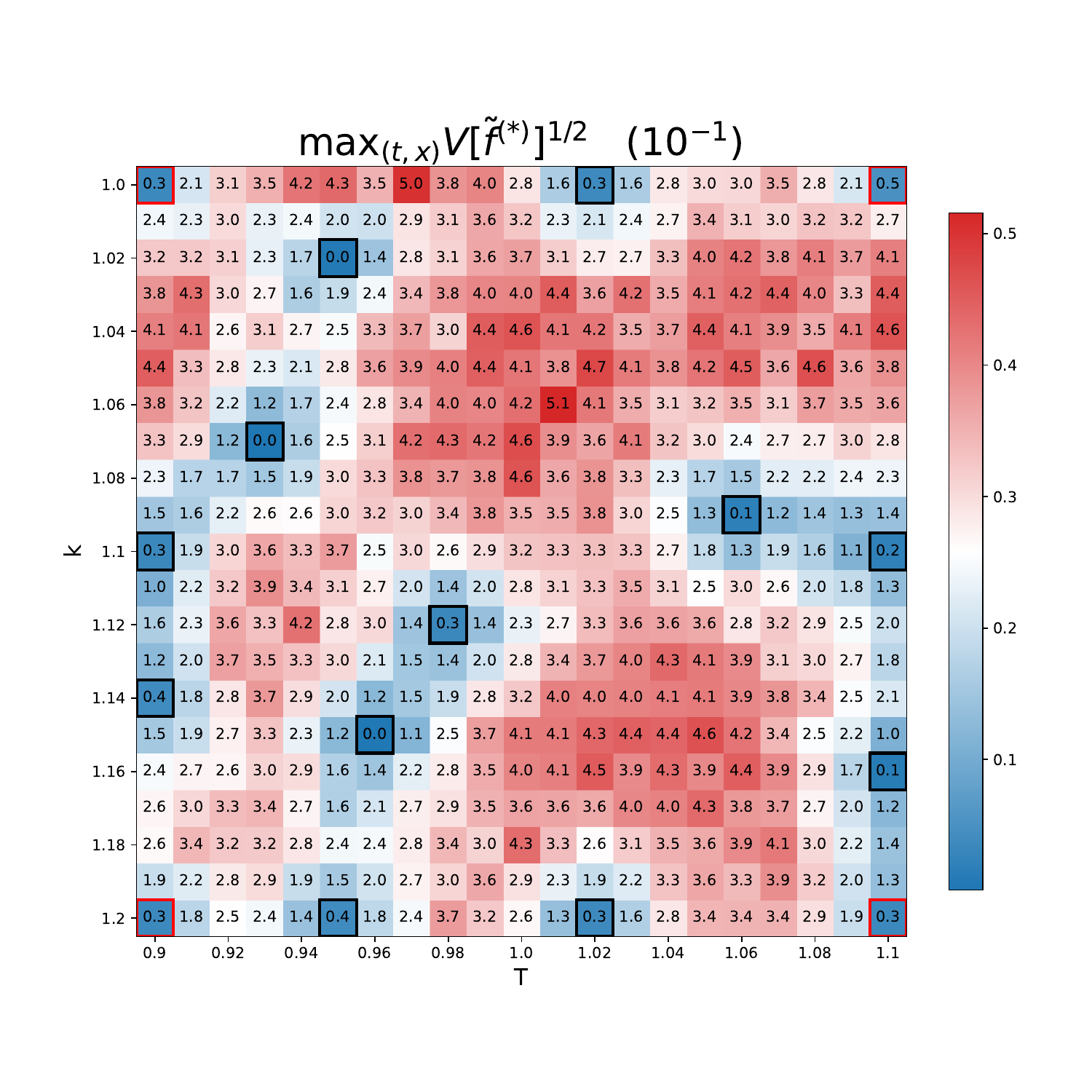}
    \caption{1D1V Vlasov Equation -- Maximum predictive standard deviation for GPLaSDI.}
    \label{vlasov_max_std}
\end{figure}

Figure \ref{vlasov_max_std} illustrates the maximum standard deviation. Although it correlates with the relative error, the correlation is only partial in this example. The standard deviation is low for parameters that correspond to a training point, indicating reproductive cases. However, the relative error can still be somewhat high in these cases. For example, for $\pmb{\mu}^{(*)}=\{0.96,1.15\}$, the maximum standard deviation $\max_{(t,x,v)}\mathbb{V}[\tilde{f}^{(*)}]^{1/2}$ is $0.0$, while the relative error $e(\tilde{f}^{(*)},f^{(*)})$ is $2.3\%$. The GPs interpolate the sets of ODEs governing the latent space dynamics, so the uncertainty quantification reflects the uncertainty in the latent space rather than the uncertainty in the training of the encoder and/or decoder. Therefore, it is possible that in some case, the uncertainty quantification in GPLaSDI may only provide a partial depiction of the model uncertainty. Using a Bayesian neural network (BNN) in place of the encoder and decoder could provide a fuller picture of the model uncertainty, but training BNNs is notoriously difficult and expensive \cite{Neal1995BayesianLF}.

\begin{figure}[!h]
\vspace{-1cm}
\hspace{-3cm}
        \includegraphics[width=1.5\textwidth]{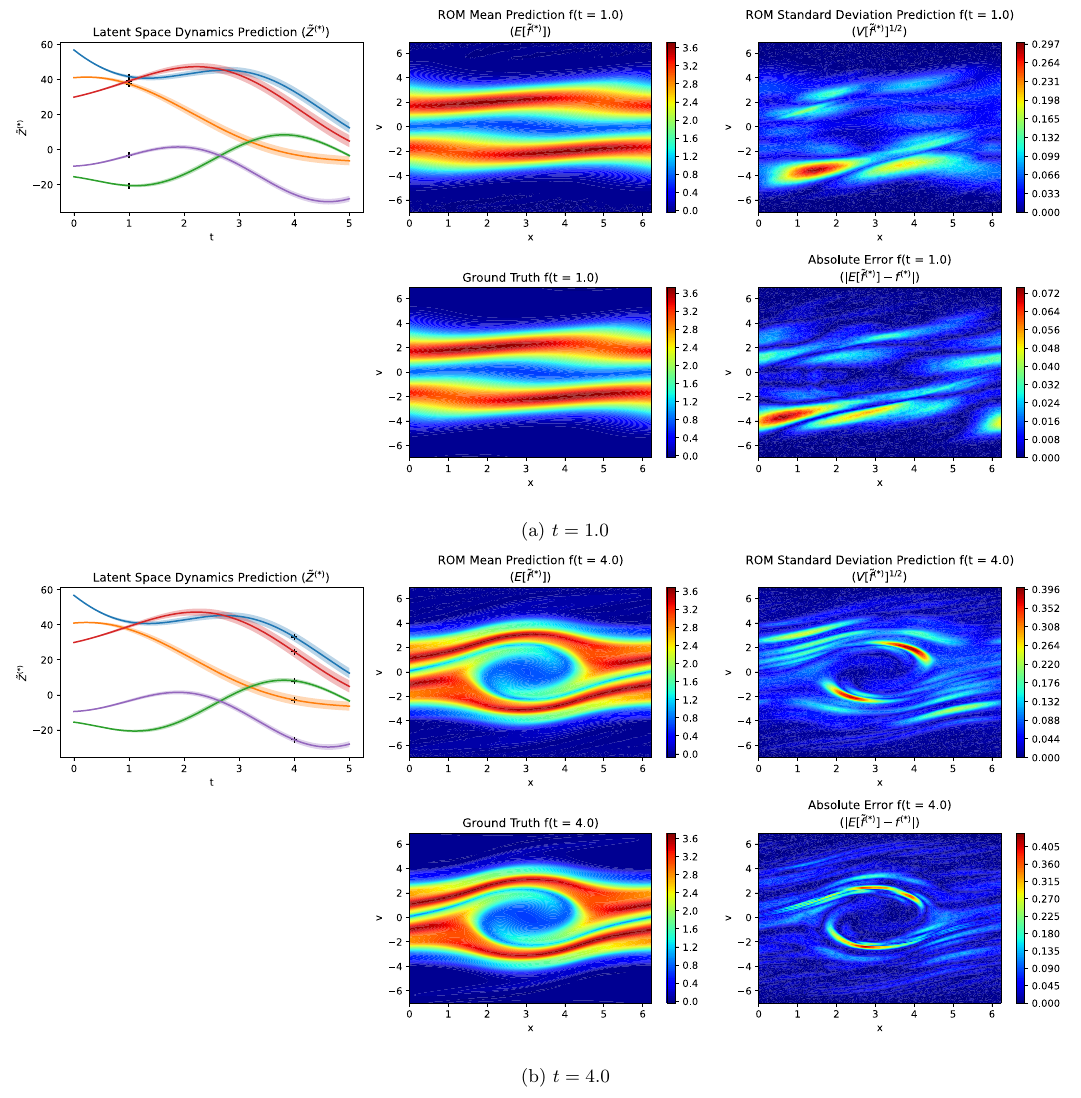}
    \caption{1D1V Vlasov Equation -- Prediction for $\pmb{\mu}^{(*)}=\{0.9,1.04\}$ at (a) $t=1.0$ and (b) $t=4.0$. These plots illustrate the predicted latent space dynamics $\mathbb{E}[\mathbf{\tilde{Z}}^{(*)}]$ along with a $95\%$ confidence interval. Additionally, the plots display the ROM mean prediction (decoder output) and standard deviation, the ground truth values, and the absolute error. 
    }
    \label{vlasov_prediction}
\end{figure}

Figure \ref{vlasov_prediction} displays the latent space dynamics, including the predicted and ground truth values of $f$, the absolute error, and the predictive standard deviation. The results correspond to the least favorable case ($\pmb{\mu}^{(*)}=\{0.9,1.04\}$) at two different time instances: $t=1$ and $t=4$. The standard deviation of the reduced-order model (ROM) exhibits qualitative similarity to the absolute error, and the error generally falls within the range of $1$ to $1.5$ standard deviations.

In $20$ separate test runs, the FOM requires an average wall clock run--time of $22.5$ seconds when utilizing four cores, and $57.9$ seconds when using a single core. In contrast, the ROM model achieves an average run--time of $1.18\cdot10^{-2}$ seconds, resulting in a remarkable average speed-up of $4906\times$ ($1906\times$ when compared to the parallel FOM). It is important to note that, similar to the Burgers equation cases, this speed-up is obtained solely using the mean prediction and does not take advantage of the full predictive distribution.

\FloatBarrier

\subsection{Rising Thermal Bubble}
\label{2D Rising Thermal Bubble}

\noindent We explore a rising thermal bubble scenario, where an initially warm bubble is introduced into a cold ambient atmosphere~\cite{ghoshconstaAIAAJ2016}. As time progresses, the bubble rises and dissipates, forming a mushroom pattern. The governing equations for this problem are the two-dimensional compressible Euler equations with gravitational source terms:
\begin{equation}
\label{euler}
\begin{cases}
    \displaystyle\frac{\partial \rho}{\partial t}+\nabla\cdot(\rho\mathbf{u})=0\hspace{0.625in}
    (t,x,y)\in[0,300]\times[0,1000]\times[0,1000]\\[10pt]
    \displaystyle\frac{\partial \rho\mathbf{u}}{\partial t}+\nabla\cdot(\rho\mathbf{u}\otimes\mathbf{u} + p\mathcal{I})=-\rho\mathbf{g}\hspace{0.3in}\mathbf{g}=\{0,9.8\}\\[10pt]
    \displaystyle\frac{\partial e}{\partial t}+\nabla\cdot\left(e+p\right){\bf u}=-\rho {\bf g}\cdot{\bf u},
\end{cases}
\end{equation}
where $\rho$ represents the fluid density, ${\bf u}=\left\{u,v\right\}$ represents the velocity, $p$ denotes the pressure, and ${\bf g}$ represents the gravitational acceleration. The internal energy $e$ is given by:
\begin{equation}
    e = \frac{p}{\gamma-1} + \frac{1}{2}\rho {\bf u} \cdot {\bf u},
\end{equation}
where $\gamma=1.4$ is the specific heat ratio. It is important to note that Equation \eqref{euler} is solved in its dimensional form. Slip-wall boundary conditions are enforced on the velocity field $\mathbf{u}$ at all boundaries. The ambient atmosphere is a hydrostatically-balanced stratified air with a constant potential temperature $\theta = 300$ and a reference pressure $p_0=10^5$. The potential temperature is defined as $\theta = T \left(\frac{p}{p_0}\right)^{\frac{\gamma}{\gamma-1}}$. A warm bubble is introduced as a potential temperature perturbation:
\begin{equation}
    \theta\Big(t=0,r=\sqrt{x^2+y^2}\Big)=
    \begin{cases}
        \frac{1}{2}\theta_c(1+\cos(\pi\frac{r}{R_c}))\hspace{0.3in}r<R_c\\
        0\hspace{1.34in}r>R_c
    \end{cases}
\end{equation}
The parameters of interest are denoted as $\pmb{\mu}=\{\theta_c,R_c\}$, representing the perturbation strength and bubble radius, respectively. The parameter $\theta_c$ ranges from $0.5$ to $0.6$, while $R_c$ ranges from $150$ to $160$. The parameter space is discretized using a $21\times21$ grid $\mathcal{D}^h$ with step sizes $\Delta \theta_c=0.005$ and $\Delta R_c=0.5$. The training process begins with $N_\mu=4$ training data points located at the four corners of the parameter space. Similarly to the Vlasov equation example, the full-order model (FOM) data is sampled using \texttt{HyPar} \cite{HyPar,ghoshconstaAIAAJ2016}. The FOM is solved using a fifth order WENO discretization~\cite{jiangshu} in space with grid spacings of $\Delta x=\Delta y = 10$, and a third order strong-stability-preserving Runge--Kutta time integration scheme with a time step size of $\Delta t=0.01$.

For the neural network architecture, we utilize a 10100--1000--200--50--20--5 configuration for the encoder (and a symmetric architecture for the decoder). The activation function employed is softplus. The latent space consists of $N_z=5$ variables, and only linear terms are considered for the SINDy library. The loss hyperparameters are set to $\beta_1=1$, $\beta_2=0.25$, and $\beta_3=10^{-6}$. To estimate the prediction variance, we use $N_s=20$ samples. The training process involves $N_{epoch}=6.8\cdot10^5$ epochs with a learning rate of $\alpha=10^{-4}$, and $N_{up}=4\cdot10^4$ updates (resulting in adding 16 data points during training, for a total of 20 training points).

\begin{figure}[!h]
\hspace{-1.5cm}
    \includegraphics[width=1.2\textwidth]{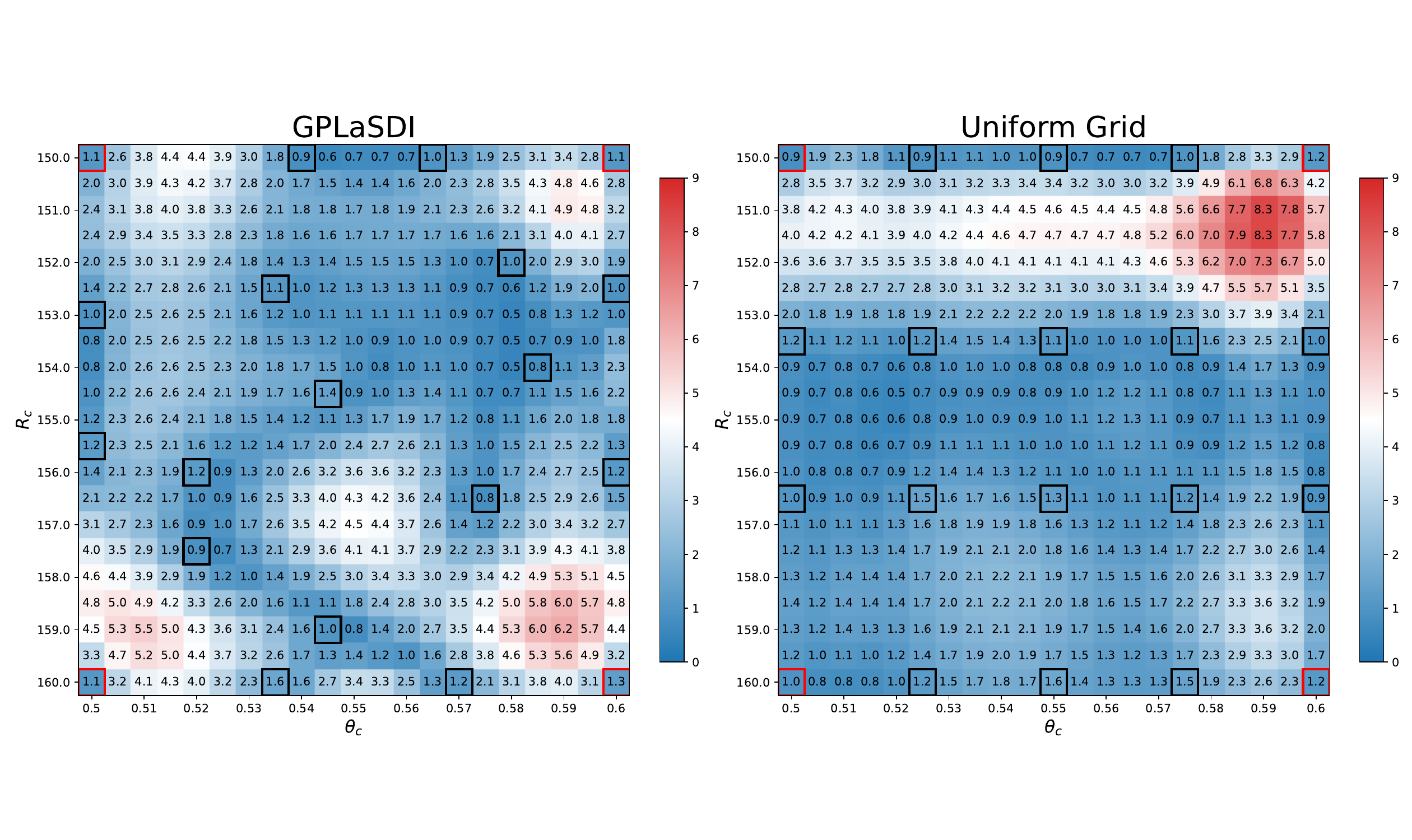}
    \caption{2D Rising Thermal Bubble -- Maximum relative error ($\%$) using GPLaSDI and a uniform training grid (non-greedy).}
    \label{rb_max_error_mean}
\end{figure}

Similar to the Vlasov equation example, we employ a baseline for comparison by training an autoencoder with the same settings and hyperparameters on a uniform parameter grid. The uniform grid consists of $20$ data points arranged in a $5\times4$ grid. Figure \ref{rb_max_error_mean} illustrates the maximum relative error for each point in the parameter space obtained using GPLaSDI and the baseline model. With GPLaSDI, the worst maximum relative error is $6.2\%$, and the largest errors occur for parameter values located towards the bottom right corner of the parameter space. GPLaSDI slightly outperforms the baseline, which exhibits higher errors for smaller values of $R_c$ ($R_c<153$), with the worst maximum relative error reaching $8.3\%$.

\begin{figure}[!h]
\hspace{2cm}
    \includegraphics[width=0.65\textwidth]{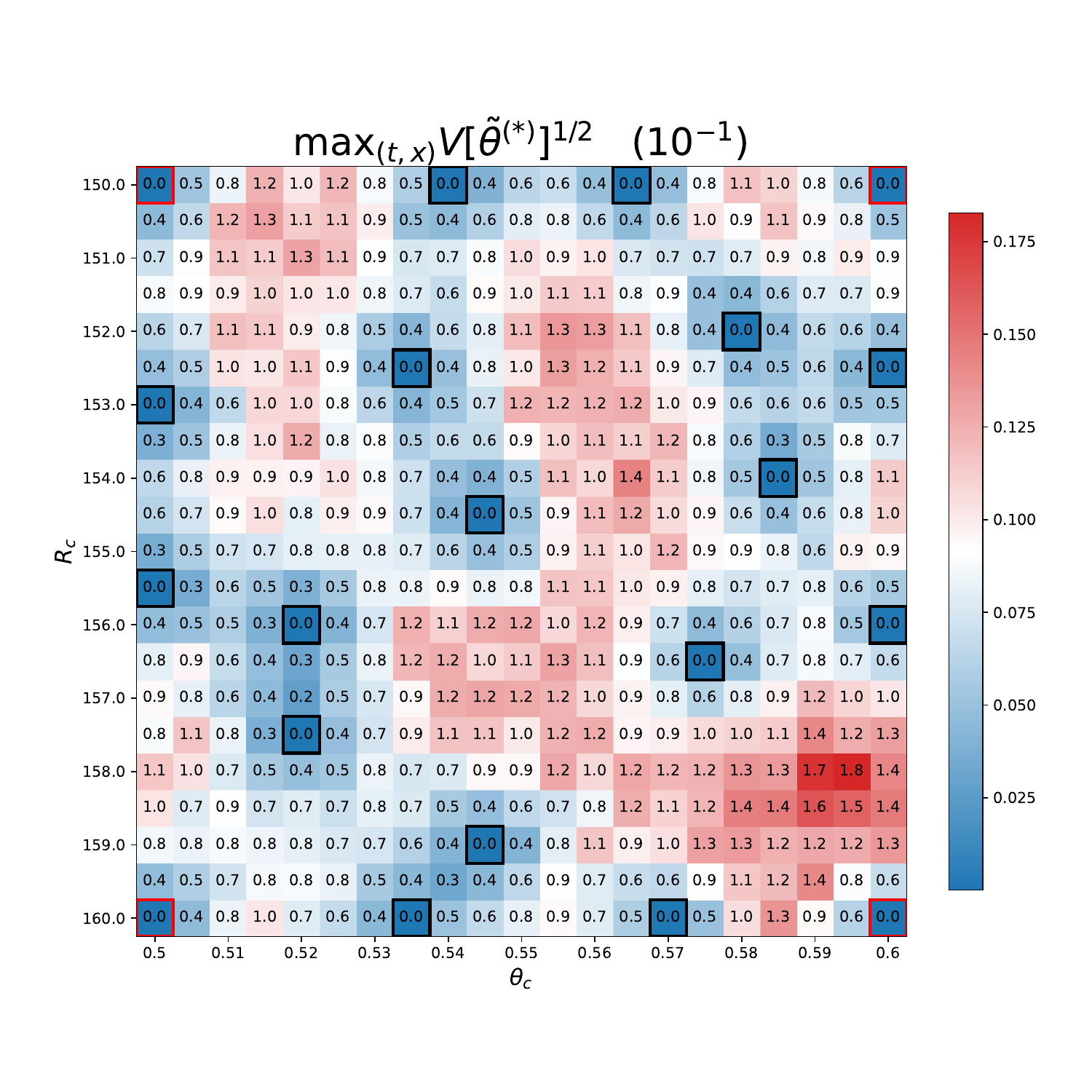}
    \caption{2D Rising Thermal Bubble -- Maximum predictive standard deviation for GPLaSDI.}
    \label{rb_max_std}
\end{figure}

Figure \ref{rb_max_std} displays the maximum standard deviation. It generally correlates reasonably well with the relative error. However, in the lower left corner of the parameter space, where large maximum relative errors are observed ($\theta_c<0.52$ and $R_c>158$), the standard deviation is unexpectedly low, erroneously indicating a high confidence in the model's predictions.

\begin{figure}[!h]
\vspace{-3cm}
    \hspace{-1.7cm}
        \includegraphics[width=1.3\textwidth]{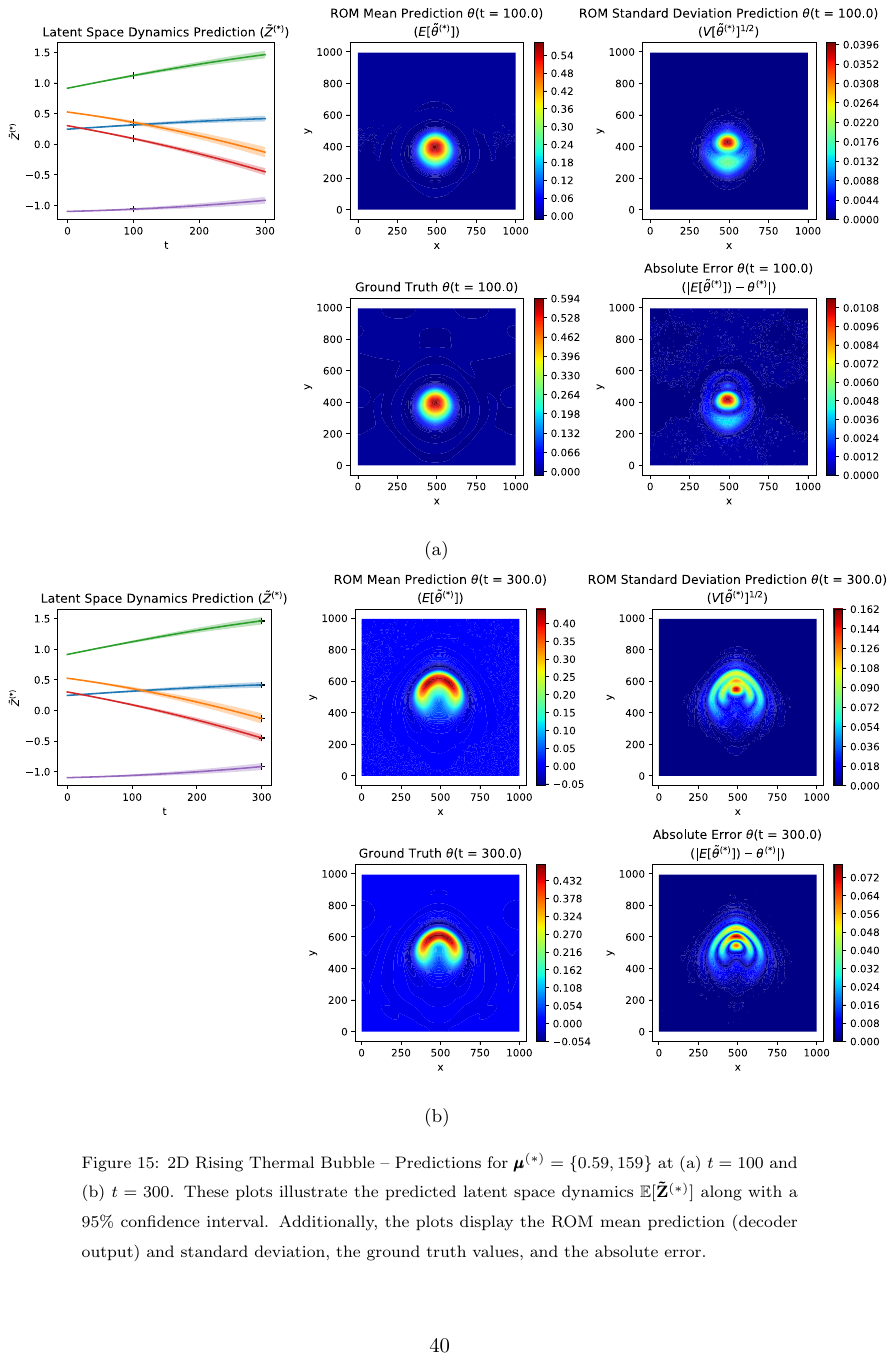}
    \caption{2D Rising Thermal Bubble -- Predictions for $\pmb{\mu}^{(*)}=\{0.59,159\}$ at (a) $t=100$ and (b) $t=300$. These plots illustrate the predicted latent space dynamics $\mathbb{E}[\mathbf{\tilde{Z}}^{(*)}]$ along with a $95\%$ confidence interval. Additionally, the plots display the ROM mean prediction (decoder output) and standard deviation, the ground truth values, and the absolute error. 
    }
    \label{rb_prediction}
\end{figure}

Figure \ref{rb_prediction} depicts the latent space dynamics, specifically the predicted and ground truth values of $\theta$, along with the absolute error and the predictive standard deviation. The results are presented for the least favorable case ($\pmb{\mu}^{(*)}=\{0.59,159\}$) at two time instances: $t=100$ and $t=300$. The absolute error typically falls within one standard deviation, and its pattern closely matches that of the standard deviation. This consistent observation aligns with the findings from the Burgers equation and Vlasov equation examples, indicating that the confidence intervals provided by GPLaSDI are meaningful and closely correlated with the prediction error.

During $20$ test runs, the FOM requires an average wall clock run-time of $89.1$ seconds when utilizing 16 cores, and $1246.8$ seconds when using a single core. On the other hand, the ROM model achieves an average run-time of $1.25\cdot10^{-2}$ seconds, resulting in an impressive average speed-up of $99744\times$ ($7128\times$ when compared to the parallel FOM). It is worth noting that, similar to the Burgers and Vlasov equation cases, this speed-up is obtained solely using the mean prediction. It is also interesting to note that throughout each example, the run time of GPLaSDI remains relatively consistent (in the order of $10^{-2}$ seconds), even though the discretization of the high-fidelity problem varied widely (and with it, the number of parameters in the autoencoder). This would indicate that GPLaSDI prediction time can scale well to problems requiring finer discretizations.

\FloatBarrier
\section{Conclusion}\label{sec:conclusion} 
We have presented GPLaSDI, a non-intrusive greedy LaSDI framework that incorporates Gaussian process latent space interpolation. Our proposed framework offers several key advantages. First, GPLaSDI efficiently captures the latent space dynamics and successfully interpolates the governing sets of ODEs while providing uncertainty quantification. This allows for meaningful confidence intervals to be obtained over the reduced-order model (ROM) predictions. These confidence intervals play a crucial role in identifying regions of high uncertainty in the parameter space. Furthermore, GPLaSDI intelligently selects additional training data points in these uncertain regions, thereby maximizing prediction accuracy while providing confidence intervals. Notably, GPLaSDI accomplishes this without requiring any prior knowledge of the underlying partial differential equation (PDE) or its residual.

We have demonstrated the effectiveness of GPLaSDI through four numerical examples, showcasing its superiority over uniform sampling baselines and its competitive performance compared to intrusive methods such as gLaSDI. GPLaSDI consistently achieved maximum relative errors of less than $6$--$7\%$, while achieving significant speed-ups ranging from several hundred to several tens of thousands of times.

Overall, GPLaSDI offers a powerful and efficient approach for capturing latent space dynamics, accurately interpolating ODEs, and providing uncertainty quantification in the context of reduced-order modeling. Its ability to autonomously select training data points and generate confidence intervals makes it a valuable tool for various scientific and engineering applications.

Currently the number of training iteration and sampling rate (and thus, the number of points that will be sampled) are all predetermined. In future work, an early termination strategy when satisfactory accuracy is obtained  could be designed. 
Additionally, the multiple GP training could become intractable in case of a large number of latent space variables (large $N_z$) and SINDy candidates (large $N_l$), since the number of ODE coefficients would grow in $\mathcal{O}(N_zN_l)$. In such case, a parallel implementation of GP training would likely be necessary.



\section*{Acknowledgements}
This research was conducted at Lawrence Livermore National Laboratory and received support from the LDRD program under project number 21-SI-006. Y. Choi also acknowledges support from the CHaRMNET Mathematical Multifaceted Integrated Capability Center (MMICC). Lawrence Livermore National Laboratory is operated by Lawrence Livermore National Security, LLC, for the U.S. Department of Energy, National Nuclear Security Administration under Contract DE-AC52-07NA27344 and LLNL-JRNL-852707.
\clearpage
\bibliography{references}

\begin{thebibliography}{10}
\expandafter\ifx\csname url\endcsname\relax
  \def\url#1{\texttt{#1}}\fi
\expandafter\ifx\csname urlprefix\endcsname\relax\def\urlprefix{URL }\fi
\expandafter\ifx\csname href\endcsname\relax
  \def\href#1#2{#2} \def\path#1{#1}\fi

\bibitem{alma991043311449403276}
S.~Raczynski, Modeling and simulation : the computer science of illusion /
  Stanislaw Raczynski., RSP Series in Computer Simulation and Modeling, John
  Wiley \& Sons, Ltd, Hertfordshire, England ;, 2006 - 2006.

\bibitem{JONES202036}
D.~Jones, C.~Snider, A.~Nassehi, J.~Yon, B.~Hicks,
  \href{https://www.sciencedirect.com/science/article/pii/S1755581720300110}{Characterising
  the digital twin: A systematic literature review}, CIRP Journal of
  Manufacturing Science and Technology 29 (2020) 36--52.
\newblock \href {https://doi.org/https://doi.org/10.1016/j.cirpj.2020.02.002}
  {\path{doi:https://doi.org/10.1016/j.cirpj.2020.02.002}}.
\newline\urlprefix\url{https://www.sciencedirect.com/science/article/pii/S1755581720300110}

\bibitem{Journal}
Review of digital twin about concepts, technologies, and industrial
  applications, Journal of Manufacturing Systems 58 (2020) 346--361.
\newblock \href {https://doi.org/10.1016/J.JMSY.2020.06.017}
  {\path{doi:10.1016/J.JMSY.2020.06.017}}.

\bibitem{article}
M.~Calder, C.~Craig, D.~Culley, R.~Cani, C.~Donnelly, R.~Douglas, B.~Edmonds,
  J.~Gascoigne, N.~Gilbert, C.~Hargrove, D.~Hinds, D.~Lane, D.~Mitchell,
  G.~Pavey, D.~Robertson, B.~Rosewell, S.~Sherwin, M.~Walport, A.~Wilson,
  Computational modelling for decision-making: Where, why, what, who and how,
  Royal Society Open Science 5 (2018) 172096.
\newblock \href {https://doi.org/10.1098/rsos.172096}
  {\path{doi:10.1098/rsos.172096}}.

\bibitem{sep-simulations-science}
E.~Winsberg, {Computer Simulations in Science}, in: E.~N. Zalta, U.~Nodelman
  (Eds.), The {Stanford} Encyclopedia of Philosophy, {W}inter 2022 Edition,
  Metaphysics Research Lab, Stanford University, 2022.

\bibitem{cummings_mason_morton_mcdaniel_2015}
R.~M. Cummings, W.~H. Mason, S.~A. Morton, D.~R. McDaniel, Applied
  Computational Aerodynamics: A Modern Engineering Approach, Cambridge
  Aerospace Series, Cambridge University Press, 2015.
\newblock \href {https://doi.org/10.1017/CBO9781107284166}
  {\path{doi:10.1017/CBO9781107284166}}.

\bibitem{6db924dfeff44d159ab577c1aefed6ef}
D.~Diston, Computational Modelling and Simulation of Aircraft and the
  Environment: Platform Kinematics and Synthetic Environment, 1st Edition,
  Vol.~1 of Aerospace Series, John Wiley \& Sons Ltd, United Kingdom, 2009.
\newblock \href {https://doi.org/10.1002/9780470744130}
  {\path{doi:10.1002/9780470744130}}.

\bibitem{car1}
K.~Kurec, M.~Remer, J.~Broniszewski, P.~Bibik, S.~Tudruj, J.~Piechna, Advanced
  modeling and simulation of vehicle active aerodynamic safety, Journal of
  Advanced Transportation 2019 (2019) 1--17.
\newblock \href {https://doi.org/10.1155/2019/7308590}
  {\path{doi:10.1155/2019/7308590}}.

\bibitem{9043275}
A.~Muhammad, I.~H. Shanono, Simulation of a car crash using ansys, in: 2019
  15th International Conference on Electronics, Computer and Computation
  (ICECCO), 2019, pp. 1--5.
\newblock \href {https://doi.org/10.1109/ICECCO48375.2019.9043275}
  {\path{doi:10.1109/ICECCO48375.2019.9043275}}.

\bibitem{Peterson_b1998}
A.~F. Peterson, S.~L. Ray, R.~Mittra.

\bibitem{rylander}
A.~B. Thomas~Rylander, Par~Ingelström, Computational Electromagnetics,
  Springer, 2013.

\bibitem{thijssen_2007}
J.~Thijssen, Computational Physics, 2nd Edition, Cambridge University Press,
  2007.
\newblock \href {https://doi.org/10.1017/CBO9781139171397}
  {\path{doi:10.1017/CBO9781139171397}}.

\bibitem{russel}
R.~Schwartz, Biological Modeling and Simulation, MIT Press, 2008.

\bibitem{10754/656260}
L.~Biegler, G.~Biros, O.~Ghattas, M.~Heinkenschloss, D.~Keyes, B.~Mallick,
  Y.~Marzouk, L.~Tenorio, B.~van Bloemen~Waanders, K.~Willcox,
  \href{http://hdl.handle.net/10754/656260}{Large-scale inverse problems and
  quantification of uncertainty} (2010).
\newblock \href {https://doi.org/10.1002/9780470685853}
  {\path{doi:10.1002/9780470685853}}.
\newline\urlprefix\url{http://hdl.handle.net/10754/656260}

\bibitem{Smith2013UncertaintyQ}
R.~C. Smith, Uncertainty quantification - theory, implementation, and
  applications, in: Computational science and engineering, 2013.

\bibitem{Sternfels_2013}
R.~Sternfels, C.~J. Earls,
  \href{https://dx.doi.org/10.1088/0266-5611/29/7/075014}{Reduced-order model
  tracking and interpolation to solve pde-based bayesian inverse problems},
  Inverse Problems 29~(7) (2013) 075014.
\newblock \href {https://doi.org/10.1088/0266-5611/29/7/075014}
  {\path{doi:10.1088/0266-5611/29/7/075014}}.
\newline\urlprefix\url{https://dx.doi.org/10.1088/0266-5611/29/7/075014}

\bibitem{https://doi.org/10.1002/nme.2746}
D.~Galbally, K.~Fidkowski, K.~Willcox, O.~Ghattas,
  \href{https://onlinelibrary.wiley.com/doi/abs/10.1002/nme.2746}{Non-linear
  model reduction for uncertainty quantification in large-scale inverse
  problems}, International Journal for Numerical Methods in Engineering 81~(12)
  (2010) 1581--1608.
\newblock \href
  {http://arxiv.org/abs/https://onlinelibrary.wiley.com/doi/pdf/10.1002/nme.2746}
  {\path{arXiv:https://onlinelibrary.wiley.com/doi/pdf/10.1002/nme.2746}},
  \href {https://doi.org/https://doi.org/10.1002/nme.2746}
  {\path{doi:https://doi.org/10.1002/nme.2746}}.
\newline\urlprefix\url{https://onlinelibrary.wiley.com/doi/abs/10.1002/nme.2746}

\bibitem{https://doi.org/10.1002/2016RS005998}
V.~Fountoulakis, C.~Earls,
  \href{https://agupubs.onlinelibrary.wiley.com/doi/abs/10.1002/2016RS005998}{Duct
  heights inferred from radar sea clutter using proper orthogonal bases}, Radio
  Science 51~(10) (2016) 1614--1626.
\newblock \href
  {http://arxiv.org/abs/https://agupubs.onlinelibrary.wiley.com/doi/pdf/10.1002/2016RS005998}
  {\path{arXiv:https://agupubs.onlinelibrary.wiley.com/doi/pdf/10.1002/2016RS005998}},
  \href {https://doi.org/https://doi.org/10.1002/2016RS005998}
  {\path{doi:https://doi.org/10.1002/2016RS005998}}.
\newline\urlprefix\url{https://agupubs.onlinelibrary.wiley.com/doi/abs/10.1002/2016RS005998}

\bibitem{do1}
S.~Wang, E.~d. Sturler, G.~H. Paulino,
  \href{https://onlinelibrary.wiley.com/doi/abs/10.1002/nme.1798}{Large-scale
  topology optimization using preconditioned krylov subspace methods with
  recycling}, International Journal for Numerical Methods in Engineering
  69~(12) (2007) 2441--2468.
\newblock \href
  {http://arxiv.org/abs/https://onlinelibrary.wiley.com/doi/pdf/10.1002/nme.1798}
  {\path{arXiv:https://onlinelibrary.wiley.com/doi/pdf/10.1002/nme.1798}},
  \href {https://doi.org/https://doi.org/10.1002/nme.1798}
  {\path{doi:https://doi.org/10.1002/nme.1798}}.
\newline\urlprefix\url{https://onlinelibrary.wiley.com/doi/abs/10.1002/nme.1798}

\bibitem{do2}
D.~White, Y.~Choi, J.~Kudo, A dual mesh method with adaptivity for
  stress-constrained topology optimization, Structural and Multidisciplinary
  Optimization 61 (02 2020).
\newblock \href {https://doi.org/10.1007/s00158-019-02393-6}
  {\path{doi:10.1007/s00158-019-02393-6}}.

\bibitem{oc}
Y.~Choi, C.~Farhat, W.~Murray, M.~Saunders,
  \href{https://arxiv.org/abs/1312.5653}{A practical factorization of a schur
  complement for pde-constrained distributed optimal control} (2013).
\newblock \href {https://doi.org/10.48550/ARXIV.1312.5653}
  {\path{doi:10.48550/ARXIV.1312.5653}}.
\newline\urlprefix\url{https://arxiv.org/abs/1312.5653}

\bibitem{doi:10.1146/annurev.fl.25.010193.002543}
G.~Berkooz, P.~Holmes, J.~L. Lumley,
  \href{https://doi.org/10.1146/annurev.fl.25.010193.002543}{The proper
  orthogonal decomposition in the analysis of turbulent flows}, Annual Review
  of Fluid Mechanics 25~(1) (1993) 539--575.
\newblock \href
  {http://arxiv.org/abs/https://doi.org/10.1146/annurev.fl.25.010193.002543}
  {\path{arXiv:https://doi.org/10.1146/annurev.fl.25.010193.002543}}, \href
  {https://doi.org/10.1146/annurev.fl.25.010193.002543}
  {\path{doi:10.1146/annurev.fl.25.010193.002543}}.
\newline\urlprefix\url{https://doi.org/10.1146/annurev.fl.25.010193.002543}

\bibitem{rbm}
G.~Rozza, D.~Huynh, A.~Patera, Reduced basis approximation and a posteriori
  error estimation for affinely parametrized elliptic coercive partial
  differential equations, Archives of Computational Methods in Engineering 15
  (2007) 1--47.
\newblock \href {https://doi.org/10.1007/BF03024948}
  {\path{doi:10.1007/BF03024948}}.

\bibitem{Safonov1988ASM}
M.~G. Safonov, R.~Y. Chiang, A schur method for balanced model reduction, 1988
  American Control Conference (1988) 1036--1040.

\bibitem{https://doi.org/10.48550/arxiv.2203.16494}
J.~T. Lauzon, S.~W. Cheung, Y.~Shin, Y.~Choi, D.~M. Copeland, K.~Huynh,
  \href{https://arxiv.org/abs/2203.16494}{S-opt: A points selection algorithm
  for hyper-reduction in reduced order models} (2022).
\newblock \href {https://doi.org/10.48550/ARXIV.2203.16494}
  {\path{doi:10.48550/ARXIV.2203.16494}}.
\newline\urlprefix\url{https://arxiv.org/abs/2203.16494}

\bibitem{doi:10.1137/19M1242963}
Y.~Choi, D.~Coombs, R.~Anderson, \href{https://doi.org/10.1137/19M1242963}{Sns:
  A solution-based nonlinear subspace method for time-dependent model order
  reduction}, SIAM Journal on Scientific Computing 42~(2) (2020) A1116--A1146.
\newblock \href {http://arxiv.org/abs/https://doi.org/10.1137/19M1242963}
  {\path{arXiv:https://doi.org/10.1137/19M1242963}}, \href
  {https://doi.org/10.1137/19M1242963} {\path{doi:10.1137/19M1242963}}.
\newline\urlprefix\url{https://doi.org/10.1137/19M1242963}

\bibitem{Stabile_2018}
G.~Stabile, G.~Rozza,
  \href{https://doi.org/10.1016%2Fj.compfluid.2018.01.035}{Finite volume
  {POD}-galerkin stabilised reduced order methods for the parametrised
  incompressible navier{\textendash}stokes equations}, Computers \& Fluids 173
  (2018) 273--284.
\newblock \href {https://doi.org/10.1016/j.compfluid.2018.01.035}
  {\path{doi:10.1016/j.compfluid.2018.01.035}}.
\newline\urlprefix\url{https://doi.org/10.1016%2Fj.compfluid.2018.01.035}

\bibitem{https://doi.org/10.1002/num.21835}
T.~Iliescu, Z.~Wang,
  \href{https://onlinelibrary.wiley.com/doi/abs/10.1002/num.21835}{Variational
  multiscale proper orthogonal decomposition: Navier-stokes equations},
  Numerical Methods for Partial Differential Equations 30~(2) (2014) 641--663.
\newblock \href
  {http://arxiv.org/abs/https://onlinelibrary.wiley.com/doi/pdf/10.1002/num.21835}
  {\path{arXiv:https://onlinelibrary.wiley.com/doi/pdf/10.1002/num.21835}},
  \href {https://doi.org/https://doi.org/10.1002/num.21835}
  {\path{doi:https://doi.org/10.1002/num.21835}}.
\newline\urlprefix\url{https://onlinelibrary.wiley.com/doi/abs/10.1002/num.21835}

\bibitem{Copeland_2022}
D.~M. Copeland, S.~W. Cheung, K.~Huynh, Y.~Choi,
  \href{https://doi.org/10.1016%2Fj.cma.2021.114259}{Reduced order models for
  lagrangian hydrodynamics}, Computer Methods in Applied Mechanics and
  Engineering 388 (2022) 114259.
\newblock \href {https://doi.org/10.1016/j.cma.2021.114259}
  {\path{doi:10.1016/j.cma.2021.114259}}.
\newline\urlprefix\url{https://doi.org/10.1016%2Fj.cma.2021.114259}

\bibitem{https://doi.org/10.48550/arxiv.2201.07335}
S.~W. Cheung, Y.~Choi, D.~M. Copeland, K.~Huynh,
  \href{https://arxiv.org/abs/2201.07335}{Local lagrangian reduced-order
  modeling for rayleigh-taylor instability by solution manifold decomposition}
  (2022).
\newblock \href {https://doi.org/10.48550/ARXIV.2201.07335}
  {\path{doi:10.48550/ARXIV.2201.07335}}.
\newline\urlprefix\url{https://arxiv.org/abs/2201.07335}

\bibitem{MCLAUGHLIN20162407}
B.~McLaughlin, J.~Peterson, M.~Ye,
  \href{https://www.sciencedirect.com/science/article/pii/S0898122116300281}{Stabilized
  reduced order models for the advection–diffusion–reaction equation using
  operator splitting}, Computers \& Mathematics with Applications 71~(11)
  (2016) 2407--2420, proceedings of the conference on Advances in Scientific
  Computing and Applied Mathematics. A special issue in honor of Max
  Gunzburger’s 70th birthday.
\newblock \href {https://doi.org/https://doi.org/10.1016/j.camwa.2016.01.032}
  {\path{doi:https://doi.org/10.1016/j.camwa.2016.01.032}}.
\newline\urlprefix\url{https://www.sciencedirect.com/science/article/pii/S0898122116300281}

\bibitem{math9141690}
Y.~Kim, K.~Wang, Y.~Choi,
  \href{https://www.mdpi.com/2227-7390/9/14/1690}{Efficient space–time
  reduced order model for linear dynamical systems in python using less than
  120 lines of code}, Mathematics 9~(14) (2021).
\newblock \href {https://doi.org/10.3390/math9141690}
  {\path{doi:10.3390/math9141690}}.
\newline\urlprefix\url{https://www.mdpi.com/2227-7390/9/14/1690}

\bibitem{https://doi.org/10.48550/arxiv.1909.11320}
Y.~Choi, G.~Oxberry, D.~White, T.~Kirchdoerfer,
  \href{https://arxiv.org/abs/1909.11320}{Accelerating design optimization
  using reduced order models} (2019).
\newblock \href {https://doi.org/10.48550/ARXIV.1909.11320}
  {\path{doi:10.48550/ARXIV.1909.11320}}.
\newline\urlprefix\url{https://arxiv.org/abs/1909.11320}

\bibitem{mcbanechoi}
S.~McBane, Y.~Choi, Component-wise reduced order model lattice-type structure
  design, Computer Methods in Applied Mechanics and Engineering 381 (2021)
  113813.
\newblock \href {https://doi.org/10.1016/j.cma.2021.113813}
  {\path{doi:10.1016/j.cma.2021.113813}}.

\bibitem{https://doi.org/10.48550/arxiv.2009.11990}
Y.~Kim, Y.~Choi, D.~Widemann, T.~Zohdi,
  \href{https://arxiv.org/abs/2009.11990}{A fast and accurate physics-informed
  neural network reduced order model with shallow masked autoencoder} (2020).
\newblock \href {https://doi.org/10.48550/ARXIV.2009.11990}
  {\path{doi:10.48550/ARXIV.2009.11990}}.
\newline\urlprefix\url{https://arxiv.org/abs/2009.11990}

\bibitem{https://doi.org/10.48550/arxiv.2011.07727}
Y.~Kim, Y.~Choi, D.~Widemann, T.~Zohdi,
  \href{https://arxiv.org/abs/2011.07727}{Efficient nonlinear manifold reduced
  order model} (2020).
\newblock \href {https://doi.org/10.48550/ARXIV.2011.07727}
  {\path{doi:10.48550/ARXIV.2011.07727}}.
\newline\urlprefix\url{https://arxiv.org/abs/2011.07727}

\bibitem{LEE2020108973}
K.~Lee, K.~T. Carlberg,
  \href{https://www.sciencedirect.com/science/article/pii/S0021999119306783}{Model
  reduction of dynamical systems on nonlinear manifolds using deep
  convolutional autoencoders}, Journal of Computational Physics 404 (2020)
  108973.
\newblock \href {https://doi.org/https://doi.org/10.1016/j.jcp.2019.108973}
  {\path{doi:https://doi.org/10.1016/j.jcp.2019.108973}}.
\newline\urlprefix\url{https://www.sciencedirect.com/science/article/pii/S0021999119306783}

\bibitem{diaz2023fast}
A.~N. Diaz, Y.~Choi, M.~Heinkenschloss, A fast and accurate
  domain-decomposition nonlinear manifold reduced order model (2023).
\newblock \href {http://arxiv.org/abs/2305.15163} {\path{arXiv:2305.15163}}.

\bibitem{doi:10.1126/science.1127647}
G.~E. Hinton, R.~R. Salakhutdinov,
  \href{https://www.science.org/doi/abs/10.1126/science.1127647}{Reducing the
  dimensionality of data with neural networks}, Science 313~(5786) (2006)
  504--507.
\newblock \href
  {http://arxiv.org/abs/https://www.science.org/doi/pdf/10.1126/science.1127647}
  {\path{arXiv:https://www.science.org/doi/pdf/10.1126/science.1127647}}, \href
  {https://doi.org/10.1126/science.1127647}
  {\path{doi:10.1126/science.1127647}}.
\newline\urlprefix\url{https://www.science.org/doi/abs/10.1126/science.1127647}

\bibitem{NIPS1992_cdc0d6e6}
D.~DeMers, G.~Cottrell,
  \href{https://proceedings.neurips.cc/paper/1992/file/cdc0d6e63aa8e41c89689f54970bb35f-Paper.pdf}{Non-linear
  dimensionality reduction}, in: S.~Hanson, J.~Cowan, C.~Giles (Eds.), Advances
  in Neural Information Processing Systems, Vol.~5, Morgan-Kaufmann, 1992.
\newline\urlprefix\url{https://proceedings.neurips.cc/paper/1992/file/cdc0d6e63aa8e41c89689f54970bb35f-Paper.pdf}

\bibitem{lasdi}
W.~D. Fries, X.~He, Y.~Choi,
  \href{https://doi.org/10.1016%2Fj.cma.2022.115436}{{LaSDI}: Parametric latent
  space dynamics identification}, Computer Methods in Applied Mechanics and
  Engineering 399 (2022) 115436.
\newblock \href {https://doi.org/10.1016/j.cma.2022.115436}
  {\path{doi:10.1016/j.cma.2022.115436}}.
\newline\urlprefix\url{https://doi.org/10.1016%2Fj.cma.2022.115436}

\bibitem{glasdi}
X.~He, Y.~Choi, W.~D. Fries, J.~Belof, J.-S. Chen,
  \href{https://www.sciencedirect.com/science/article/abs/pii/S0021999123003625}{{gLaSDI}:
  Parametric physics-informed greedy latent space dynamics identification} 489
  (2023) 112267.
\newblock \href {https://doi.org/10.1016/j.jcp.2023.112267}
  {\path{doi:10.1016/j.jcp.2023.112267}}.
\newline\urlprefix\url{https://www.sciencedirect.com/science/article/abs/pii/S0021999123003625}

\bibitem{osti_1420279}
G.~Tapia, S.~A. Khairallah, M.~J. Matthews, W.~E. King, A.~Elwany, Gaussian
  process-based surrogate modeling framework for process planning in laser
  powder-bed fusion additive manufacturing of 316l stainless steel,
  International Journal of Advanced Manufacturing Technology 94~(9-12) (9
  2017).
\newblock \href {https://doi.org/10.1007/s00170-017-1045-z}
  {\path{doi:10.1007/s00170-017-1045-z}}.

\bibitem{Marjavaara2006CFDDO}
D.~Marjavaara, Cfd driven optimization of hydraulic turbine draft tubes using
  surrogate models, 2006.

\bibitem{unknown}
K.~Cheng, R.~Zimmermann, Sliced gradient-enhanced kriging for high-dimensional
  function approximation and aerodynamic modeling (04 2022).

\bibitem{kutz_2017}
J.~N. Kutz, Deep learning in fluid dynamics, Journal of Fluid Mechanics 814
  (2017) 1–4.
\newblock \href {https://doi.org/10.1017/jfm.2016.803}
  {\path{doi:10.1017/jfm.2016.803}}.

\bibitem{koza:1994:SandC}
J.~R. Koza, Genetic programming as a means for programming computers by natural
  selection, Statistics and Computing 4~(2) (1994) 87--112.
\newblock \href {https://doi.org/doi:10.1007/BF00175355}
  {\path{doi:doi:10.1007/BF00175355}}.

\bibitem{doi:10.1126/science.1165893}
M.~Schmidt, H.~Lipson,
  \href{https://www.science.org/doi/abs/10.1126/science.1165893}{Distilling
  free-form natural laws from experimental data}, Science 324~(5923) (2009)
  81--85.
\newblock \href
  {http://arxiv.org/abs/https://www.science.org/doi/pdf/10.1126/science.1165893}
  {\path{arXiv:https://www.science.org/doi/pdf/10.1126/science.1165893}}, \href
  {https://doi.org/10.1126/science.1165893}
  {\path{doi:10.1126/science.1165893}}.
\newline\urlprefix\url{https://www.science.org/doi/abs/10.1126/science.1165893}

\bibitem{doi:10.1073/pnas.1517384113}
S.~L. Brunton, J.~L. Proctor, J.~N. Kutz,
  \href{https://www.pnas.org/doi/abs/10.1073/pnas.1517384113}{Discovering
  governing equations from data by sparse identification of nonlinear dynamical
  systems}, Proceedings of the National Academy of Sciences 113~(15) (2016)
  3932--3937.
\newblock \href
  {http://arxiv.org/abs/https://www.pnas.org/doi/pdf/10.1073/pnas.1517384113}
  {\path{arXiv:https://www.pnas.org/doi/pdf/10.1073/pnas.1517384113}}, \href
  {https://doi.org/10.1073/pnas.1517384113}
  {\path{doi:10.1073/pnas.1517384113}}.
\newline\urlprefix\url{https://www.pnas.org/doi/abs/10.1073/pnas.1517384113}

\bibitem{doi:10.1126/sciadv.1602614}
S.~H. Rudy, S.~L. Brunton, J.~L. Proctor, J.~N. Kutz,
  \href{https://www.science.org/doi/abs/10.1126/sciadv.1602614}{Data-driven
  discovery of partial differential equations}, Science Advances 3~(4) (2017)
  e1602614.
\newblock \href
  {http://arxiv.org/abs/https://www.science.org/doi/pdf/10.1126/sciadv.1602614}
  {\path{arXiv:https://www.science.org/doi/pdf/10.1126/sciadv.1602614}}, \href
  {https://doi.org/10.1126/sciadv.1602614} {\path{doi:10.1126/sciadv.1602614}}.
\newline\urlprefix\url{https://www.science.org/doi/abs/10.1126/sciadv.1602614}

\bibitem{https://doi.org/10.48550/arxiv.2211.10575}
L.~M. Gao, J.~N. Kutz, \href{https://arxiv.org/abs/2211.10575}{Bayesian
  autoencoders for data-driven discovery of coordinates, governing equations
  and fundamental constants} (2022).
\newblock \href {https://doi.org/10.48550/ARXIV.2211.10575}
  {\path{doi:10.48550/ARXIV.2211.10575}}.
\newline\urlprefix\url{https://arxiv.org/abs/2211.10575}

\bibitem{https://doi.org/10.48550/arxiv.2205.10965}
K.~Owens, J.~N. Kutz, \href{https://arxiv.org/abs/2205.10965}{Data-driven
  discovery of governing equations for coarse-grained heterogeneous network
  dynamics} (2022).
\newblock \href {https://doi.org/10.48550/ARXIV.2205.10965}
  {\path{doi:10.48550/ARXIV.2205.10965}}.
\newline\urlprefix\url{https://arxiv.org/abs/2205.10965}

\bibitem{doi:10.1098/rsos.211823}
S.~M. Hirsh, D.~A. Barajas-Solano, J.~N. Kutz,
  \href{https://royalsocietypublishing.org/doi/abs/10.1098/rsos.211823}{Sparsifying
  priors for bayesian uncertainty quantification in model discovery}, Royal
  Society Open Science 9~(2) (2022) 211823.
\newblock \href
  {http://arxiv.org/abs/https://royalsocietypublishing.org/doi/pdf/10.1098/rsos.211823}
  {\path{arXiv:https://royalsocietypublishing.org/doi/pdf/10.1098/rsos.211823}},
  \href {https://doi.org/10.1098/rsos.211823} {\path{doi:10.1098/rsos.211823}}.
\newline\urlprefix\url{https://royalsocietypublishing.org/doi/abs/10.1098/rsos.211823}

\bibitem{Messenger_2021}
D.~A. Messenger, D.~M. Bortz,
  \href{https://doi.org/10.1016%2Fj.jcp.2021.110525}{Weak {SINDy} for partial
  differential equations}, Journal of Computational Physics 443 (2021) 110525.
\newblock \href {https://doi.org/10.1016/j.jcp.2021.110525}
  {\path{doi:10.1016/j.jcp.2021.110525}}.
\newline\urlprefix\url{https://doi.org/10.1016%2Fj.jcp.2021.110525}

\bibitem{Chen_2021}
Z.~Chen, Y.~Liu, H.~Sun,
  \href{https://doi.org/10.1038%2Fs41467-021-26434-1}{Physics-informed learning
  of governing equations from scarce data}, Nature Communications 12~(1) (oct
  2021).
\newblock \href {https://doi.org/10.1038/s41467-021-26434-1}
  {\path{doi:10.1038/s41467-021-26434-1}}.
\newline\urlprefix\url{https://doi.org/10.1038%2Fs41467-021-26434-1}

\bibitem{BONNEVILLE2022100115}
C.~Bonneville, C.~Earls,
  \href{https://www.sciencedirect.com/science/article/pii/S2590055222000117}{Bayesian
  deep learning for partial differential equation parameter discovery with
  sparse and noisy data}, Journal of Computational Physics: X 16 (2022) 100115.
\newblock \href {https://doi.org/https://doi.org/10.1016/j.jcpx.2022.100115}
  {\path{doi:https://doi.org/10.1016/j.jcpx.2022.100115}}.
\newline\urlprefix\url{https://www.sciencedirect.com/science/article/pii/S2590055222000117}

\bibitem{STEPHANY2022360}
R.~Stephany, C.~Earls,
  \href{https://www.sciencedirect.com/science/article/pii/S0893608022002660}{Pde-read:
  Human-readable partial differential equation discovery using deep learning},
  Neural Networks 154 (2022) 360--382.
\newblock \href {https://doi.org/https://doi.org/10.1016/j.neunet.2022.07.008}
  {\path{doi:https://doi.org/10.1016/j.neunet.2022.07.008}}.
\newline\urlprefix\url{https://www.sciencedirect.com/science/article/pii/S0893608022002660}

\bibitem{https://doi.org/10.48550/arxiv.2212.04971}
R.~Stephany, C.~Earls, \href{https://arxiv.org/abs/2212.04971}{Pde-learn: Using
  deep learning to discover partial differential equations from noisy, limited
  data} (2022).
\newblock \href {https://doi.org/10.48550/ARXIV.2212.04971}
  {\path{doi:10.48550/ARXIV.2212.04971}}.
\newline\urlprefix\url{https://arxiv.org/abs/2212.04971}

\bibitem{doi:10.1073/pnas.1906995116}
K.~Champion, B.~Lusch, J.~N. Kutz, S.~L. Brunton,
  \href{https://www.pnas.org/doi/abs/10.1073/pnas.1906995116}{Data-driven
  discovery of coordinates and governing equations}, Proceedings of the
  National Academy of Sciences 116~(45) (2019) 22445--22451.
\newblock \href
  {http://arxiv.org/abs/https://www.pnas.org/doi/pdf/10.1073/pnas.1906995116}
  {\path{arXiv:https://www.pnas.org/doi/pdf/10.1073/pnas.1906995116}}, \href
  {https://doi.org/10.1073/pnas.1906995116}
  {\path{doi:10.1073/pnas.1906995116}}.
\newline\urlprefix\url{https://www.pnas.org/doi/abs/10.1073/pnas.1906995116}

\bibitem{https://doi.org/10.48550/arxiv.2106.09658}
Z.~Bai, L.~Peng, \href{https://arxiv.org/abs/2106.09658}{Non-intrusive
  nonlinear model reduction via machine learning approximations to
  low-dimensional operators} (2021).
\newblock \href {https://doi.org/10.48550/ARXIV.2106.09658}
  {\path{doi:10.48550/ARXIV.2106.09658}}.
\newline\urlprefix\url{https://arxiv.org/abs/2106.09658}

\bibitem{books/lib/RasmussenW06}
C.~E. Rasmussen, C.~K.~I. Williams, Gaussian processes for machine learning.,
  Adaptive computation and machine learning, MIT Press, 2006.

\bibitem{Goodfellow-et-al-2016}
I.~Goodfellow, Y.~Bengio, A.~Courville, Deep Learning, MIT Press, 2016,
  \url{http://www.deeplearningbook.org}.

\bibitem{adam}
D.~P. Kingma, J.~Ba, \href{https://arxiv.org/abs/1412.6980}{Adam: A method for
  stochastic optimization} (2014).
\newblock \href {https://doi.org/10.48550/ARXIV.1412.6980}
  {\path{doi:10.48550/ARXIV.1412.6980}}.
\newline\urlprefix\url{https://arxiv.org/abs/1412.6980}

\bibitem{wilson2015kernel}
A.~G. Wilson, H.~Nickisch, Kernel interpolation for scalable structured
  gaussian processes (kiss-gp) (2015).
\newblock \href {http://arxiv.org/abs/1503.01057} {\path{arXiv:1503.01057}}.

\bibitem{wilson2015thoughts}
A.~G. Wilson, C.~Dann, H.~Nickisch, Thoughts on massively scalable gaussian
  processes (2015).
\newblock \href {http://arxiv.org/abs/1511.01870} {\path{arXiv:1511.01870}}.

\bibitem{muyskens2021muygps}
A.~Muyskens, B.~Priest, I.~Goumiri, M.~Schneider, Muygps: Scalable gaussian
  process hyperparameter estimation using local cross-validation (2021).
\newblock \href {http://arxiv.org/abs/2104.14581} {\path{arXiv:2104.14581}}.

\bibitem{bishop2007}
C.~M. Bishop,
  \href{http://www.amazon.com/Pattern-Recognition-Learning-Information-Statistics/dp/0387310738%3FSubscriptionId%3D13CT5CVB80YFWJEPWS02%26tag%3Dws%26linkCode%3Dxm2%26camp%3D2025%26creative%3D165953%26creativeASIN%3D0387310738}{Pattern
  Recognition and Machine Learning (Information Science and Statistics)}, 1st
  Edition, Springer, 2007.
\newline\urlprefix\url{http://www.amazon.com/Pattern-Recognition-Learning-Information-Statistics/dp/0387310738%3FSubscriptionId%3D13CT5CVB80YFWJEPWS02%26tag%3Dws%26linkCode%3Dxm2%26camp%3D2025%26creative%3D165953%26creativeASIN%3D0387310738}

\bibitem{NEURIPS2019_9015}
A.~Paszke, S.~Gross, F.~Massa, A.~Lerer, J.~Bradbury, G.~Chanan, T.~Killeen,
  Z.~Lin, N.~Gimelshein, L.~Antiga, A.~Desmaison, A.~Kopf, E.~Yang, Z.~DeVito,
  M.~Raison, A.~Tejani, S.~Chilamkurthy, B.~Steiner, L.~Fang, J.~Bai,
  S.~Chintala,
  \href{http://papers.neurips.cc/paper/9015-pytorch-an-imperative-style-high-performance-deep-learning-library.pdf}{Pytorch:
  An imperative style, high-performance deep learning library}, in: Advances in
  Neural Information Processing Systems 32, Curran Associates, Inc., 2019, pp.
  8024--8035.
\newline\urlprefix\url{http://papers.neurips.cc/paper/9015-pytorch-an-imperative-style-high-performance-deep-learning-library.pdf}

\bibitem{pedregosa2011scikit}
F.~Pedregosa, G.~Varoquaux, A.~Gramfort, V.~Michel, B.~Thirion, O.~Grisel,
  M.~Blondel, P.~Prettenhofer, R.~Weiss, V.~Dubourg, et~al., Scikit-learn:
  Machine learning in python, Journal of machine learning research 12~(Oct)
  (2011) 2825--2830.

\bibitem{HyPar}
{HyPar} {R}epository, {\tt https://bitbucket.org/deboghosh/hypar}.

\bibitem{jiangshu}
G.-S. Jiang, C.-W. Shu, Efficient implementation of weighted {ENO} schemes,
  Journal of Computational Physics 126~(1) (1996) 202--228.
\newblock \href {https://doi.org/10.1006/jcph.1996.0130}
  {\path{doi:10.1006/jcph.1996.0130}}.

\bibitem{Neal1995BayesianLF}
R.~M. Neal, \href{https://api.semanticscholar.org/CorpusID:60809283}{Bayesian
  learning for neural networks}, 1995.
\newline\urlprefix\url{https://api.semanticscholar.org/CorpusID:60809283}

\bibitem{ghoshconstaAIAAJ2016}
D.~Ghosh, E.~M. Constantinescu, Well-balanced, conservative finite-difference
  algorithm for atmospheric flows, AIAA Journal 54~(4) (2016) 1370--1385.
\newblock \href {https://doi.org/10.2514/1.J054580}
  {\path{doi:10.2514/1.J054580}}.

\end{thebibliography}

\end{document}